\documentclass{article}
\usepackage[utf8]{inputenc}
\usepackage[english]{babel}
\usepackage{fancyhdr}
\usepackage{dsfont} %eins als neutrales element

\usepackage{caption,chngcntr}

\usepackage{mathrsfs}
\usepackage{enumitem} 
\usepackage[ %^showframe
  left=3.7cm,    
  right=3.4cm,
  top=2.5cm,
  bottom=3.5cm,
]{geometry}
\usepackage{fancybox} 
\usepackage[T1]{fontenc} 
\usepackage[arrow, matrix, curve]{xy}
\usepackage{amsthm}
\usepackage{dsfont}
\usepackage{graphicx} 
\graphicspath{{./Abbildungen/}}  
\usepackage{tabularx}
\usepackage{amssymb} %fÃŒr doppelte Buchstaben, wie die Reellen Zahlen.
\usepackage{amsmath} %zB fÃŒr matrizen
\usepackage{color}
\usepackage{framed}
\usepackage{endnotes} % Fuer Fussnote

\usepackage[onehalfspacing]{setspace}
\usepackage[round]{natbib}
\usepackage{tikz}
\usetikzlibrary{shapes,angles,quotes}

\usetikzlibrary{calc}
\usetikzlibrary{matrix}
\usepackage{pgfplots}
\pgfplotsset{compat=1.11}
\usepackage{braket}

\theoremstyle{definition}

\newtheorem{definition}{Definition}
\newtheorem{example}{Example}

\usepackage{varioref}
\title{Sub-Riemannian geometry applied to incompressible, inviscid fluids}
\author{Annette M\"uller and Peter N\'evir}
\date{March 2022}

\begin{document}

\maketitle

\begin{figure}[t]
\centering{
\includegraphics[scale=.9]{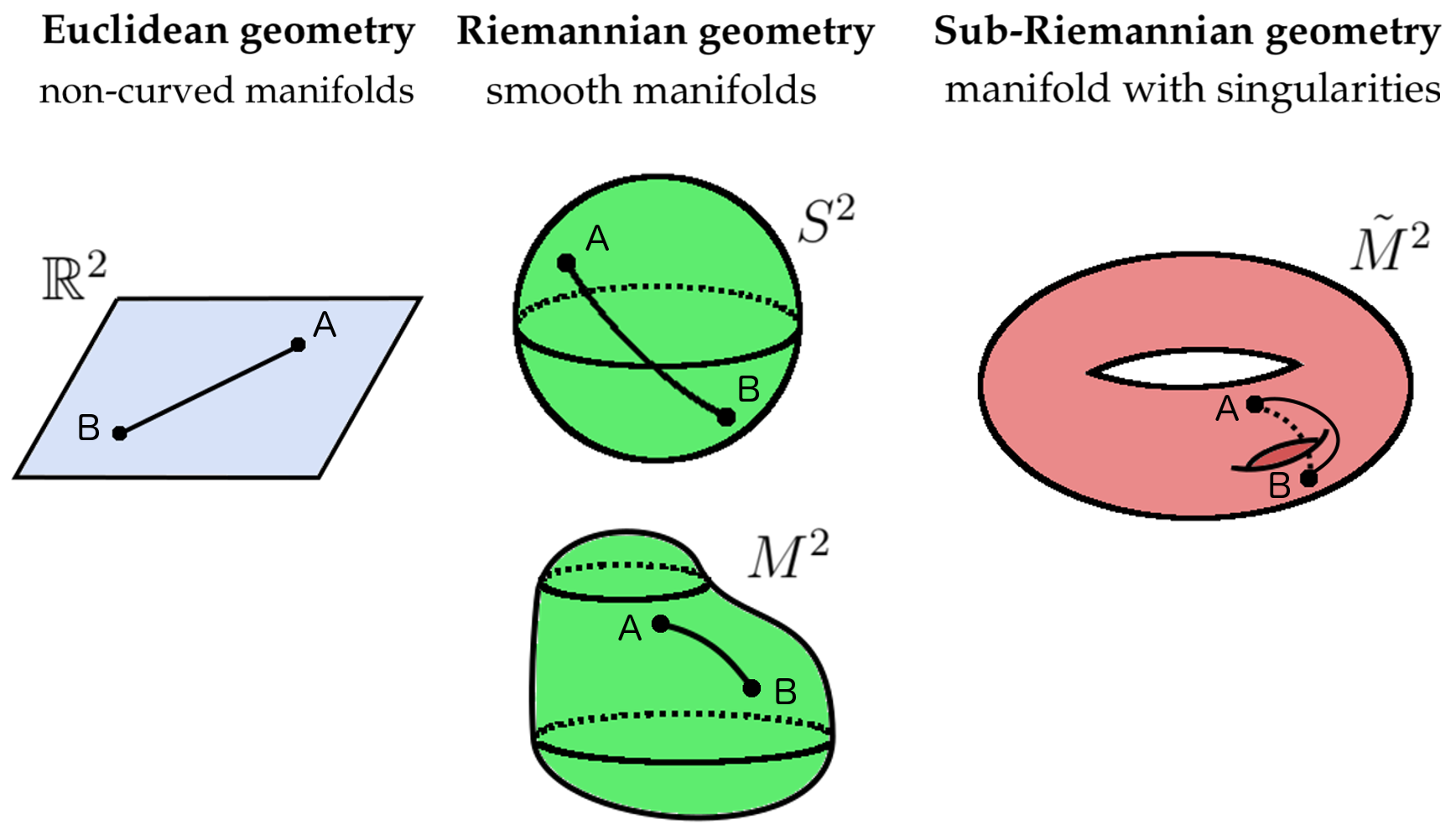} %\includegraphics[scale=.8]{2Wirbel_2D.eps}
\caption{\small{The differences between Euclidean, Riemannian and sub-Riemannian geometry are shown. All spaces can be endowed with a suitable metric to connect two points $A$ and $B$ on a manifold. In case of a non-smooth manifold, we have to move on so-called horizontal bundles, because the motion is restricted by singularities. Sub-Riemannian geometry is also called singular Riemannian geometry.}}
\label{fig:SUBR}
}
\end{figure}

\section*{Abstract}
One field of fluid dynamics concerns the search for variational principles. 
So far,
the Hamiltonian view and Riemannian geometry has been applied to find
geodesics for hydrodynamic systems. Compared to Riemannian geometry sub-Riemannian geometry can be applied to search for geodesics
of constrained systems such as vortex flows, where the vortex motion is
restricted by rotations that are expressed by vortex-related conservation laws.
The Nambu formulation of incompressible, inviscid fluid dynamics provides a nilpotent Lie algebra for two- and three-dimensional fluids, called Vortex-Heisenberg algebra, which makes it natural to apply sub-Riemannian geometry. In our first approach, we consider discretized models.
The resulting vortex geodesics for two-dimensional incompressible, inviscid  discrete point vortices is a known special point vortex constellation, the equilibrium. We also outline a concept for the derivation of 3D vortex geodesics.
 %We will show how the derived
%2D vortex geodesics correspond to the well known point vortex equilibria,

%will show how the vortex algebra allows for the investigation of shortest
%paths of point vortices. Such vortex geodesics can be compared to special
%point vortex constellations that we will have discussed in the first part of
%the thesis. We will also outline a concept for the derivation of 3D vortex geodesics....

\section{Introduction}

In the following, we will apply sub-Riemannian geometry to search for geodesics.  
But what are geodesics in terms of vortex dynamics? In a first approach, we consider the discrete, idealized point vortex model for two-dimensional incompressible, inviscid vortex motion, where vortices are compactified to points. These point vortices are characterized by their circulation, which describes the strength and the direction of rotation. The circulation is a conserved quantity on two-dimensional material surfaces, see e.g. \citet{Aref2007} or \citet{Newton2001}.
We consider vortex geodesics as shortest paths of single point vortices. 
We also outline a concept, how sub-Riemannian geometry can be applied to find shortest paths in three spatial dimensions.

%We note two differences between the fluid mechanical Lie algebras that are based on Poisson brackets as shown above and the Lie algebras based on Nambu-mechanics derived by \citet{Nevir1993}.
%On the one hand, the antisymmetric Poisson-Lie brackets yields a representation of the fluid dynamical equations in terms of the energy. %Classically, the vorticity and the velocity field are analyzed separately. 

Most algebraic formulations and derivations of variational principles for hydrodynamical systems are based on the Hamiltonian structure, where the Poisson bracket is considered
\citet{Arnold1969}, \citet{Arnold1969one}, \citet{Salmon1982}, \citet{Marsden1983}, \citet{Salmon1988}, \citet{Shepherd1990}, \citet{Arnold1992}, or \cite{Holm1998}.
The here discussed derivation of point vortex geodesics is based on the Nambu representation of the equations of incompressible, inviscid fluid dynamics. It is derived from the Helmholtz vorticity equation. In contrast to the well-known Poisson-bracket, the Nambu bracket has three arguments, where a vortex conservation law (2D: enstrophy, 3D: helicity) has equal status as the energy. 
The Nambu bracket generates a nilpotent Lie algebra for incompressiblel, inviscid fluids \citep[see][]{Nevir1993,Nevir1998}.
In \citet{muller2021algebra} this algebra is discussed in detail and a matrix representation of the Lie algebra is introduced. Moreover, the authors use the property of being nilpotent to derive corresponding Vortex-Heisenberg groups for two- and three-dimensional incompressible, inviscid fluids.
In the following, we also use the nilpotent structure that allows for the application of sub-Riemannian geometry. This work has been published in the framework of a PhD thesis in \citet{mueller2018algebraic}.

%Applying sub-Riemannian geometry to vortex dynamics we will first regard pure mathematics of the fields differential geometry and algebra. Then, we will apply this theoretical concept to fluid dynamics, which further can be transferred to atmospheric phenomena. 
In general, in order to find geodesics, one first has to find appropriate metrics to measure distances.
Mathematically, there are many different metric spaces, each defined by a set with a metric. 
Let us start start with the well-known Euclidean geometry in three dimensions. The Euclidean distance between two points is given by a straight line segment, which is the shortest path between two points and therefore the geodesic.

This concept can be generalized to find shortest paths on arbitrary smooth manifolds, as we have sketched in fig. \ref{fig:SUBR}. Let us consider a smooth curved surface $M$, for example the sphere $M=S^2$ embedded in a three dimensional space, and denote with $A$ a starting point on this curved surface $M$. We look for the shortest path from $A$ to another point $B$ on $M$. 
In this case, Euclidean geometry is not an appropriate choice to measure the distance between $A$ and $B$, because it would be the secant line; the shortest path from $A$ to $B$ on the manifold is curved, too, and can not be a segment of a straight line. We would like to move \textit{on} the surface, as for example on the sphere, and we assume that we are not allowed to cross the sphere. 
As a first guess, we would use Riemannian geometry, where the so-called Riemannian metric is defined with respect to the manifold such that the Riemannian metric can be used to measures lengths of paths on \textit{any smooth} manifold. On a sphere it turns out that the shortest path between two points lies on a great circle, see fig. \ref{fig:SUBR}, and the geodesics is given by a formula of the arccosine.

Sometimes, constraints restrict the motion such that the manifold is not smooth anymore. In this case, we have to measure distances using tangent spaces.
At each point on a Riemannian manifold the tangent space is endowed with the Euclidean structure. This structure smoothly depends on the point where the tangent space is attached. 
Let us assume, we have a walk on a Riemannian manifold $M$ and we stop at a point $p$ in $M$. Then, the tangent vectors in this point give us the directions where to move. We can move in all directions on the tangent plane. And we can measure lengths of vectors and angles between vectors that are attached at the same point. These measurements are done using the Euclidean rules. 

Let us now assume that we stand on a sub-Riemannian manifold. We are not allowed to move in all directions. There are constraints, for example a physical law, that restrict our motion. A sub-Riemannian space is a smooth manifold with a fixed admissible subspace in any tangent space where the admissible subspaces are equipped with Euclidean structures \citep{Barilari2016,Barilari2012}. %\citetCalin{2009}, \citet{Montgomery2006}
The distance between two points in a sub-Riemannian space is the infimum of the length of admissible paths connecting the points. 
As we will discuss more in detail later, sub-\-Riemannian geodesics are measured by moving along curves that are tangent to so-called horizontal subspaces.
%More precisely, sub-Riemannian geometry can be applied to find geodesics of constrained systems.  
For example, we would use sub-Riemannian geometry if we aim for finding the shortest orbits of satellites in space, or if we park a car. In the last example our constraint is given by the fact that we can not drive a car sidewards. Thus, to describe the position of a car we consider its location ($\mathbb{R}^2$) and an angle ($S^1$), i.e. a point on the manifold $\mathbb{R}^2 \times S^1$.
Its shortest path can be determined by the infimum of a sub-Riemannian path, we call this shortest path sub-Riemannian goedesics. For further readings see, e.g.,  \citet{Montgomery2006}, \citet{Calin2009} or \citet{Barilari2016}.

From mathematical perspective, nilpotent groups allow for the application of sub-Riemannian geometry. 
Therefore, the classical Heisenberg group is an example for the derivation of sub-Riemannian geo\-desics. 
Physically, the group representation for electric charged particles in static inhomogeneous magnetic fields is given by a Heisenberg group \citep{Monroy1999,Montgomery2006}. 
Classical mass points move on straight lines, therefore, we can find their shortest paths without applying sub-Riemannian geometry. But charged particles behave similar to vortices.

Here, we will consider the Vortex-Heisenberg groups for incompressible two- and three dimensional fluid dynamics VH(2) and VH(3), see \citet{muller2021algebra}.
These vortex groups hold the classical Heisenberg group structure. 
%We will call these shortest paths of vortices in incompressible, inviscid fluids shortly \textit{vortex geodesics}.
Physically, vortex motion is constrained by conservation laws: 
the conservation of the linear momentum, the angular momentum and the energy can all be expressed via the vorticity. There are also scale-dependent conservation laws for incompressible, inviscid fluids such as the enstrophy and circulation in two dimensions and the helicity and the flux of vorticity in three dimensions.
Therefore, the motion of vortices is constrained by vortical rotations. We think that the  constraints on the vortex motions are mathematically reflected in the nilpotent structure of the Vortex-Heisenberg group.
This nilpotent structure provides a natural sub-Riemannian applicability.

But, so far, sub-Riemannian geometry has not been considered for the study of vortex dynamics. Instead, the Riemannian view has been used for the investigation of extremal principles for hydrodynamic systems, see e.g. \citet{Arnold1992} or \citet{Holm1998}. where mostly the energy is considered to derive extremal principles for fluid dynamical systems. But, to the best of our knowledge, there are no investigations of the derivation of vortex geodesics regarding additional vortex-related quantities. 
%The major difference is the set of equation that provides the underlying structure. 
%The Euler equation is commonly used as basis for the geometrical as well as for the algebraical view. 
%In contrast, regarding the Nambu formulation, the algebraic and geometric views on vortex dynamics are based on the Helmholtz equation.
%This means that most authors consider wind field as basis variable and analyze the kinetic energy and the vortex quantities based on the Euler equations. 
%The Helmholtz equation results from the rotation of the Euler equations leading to the description of the time evolution of the vorticity. But, some authors do consider the vorticity equation as basis, but they still deal with the Hamiltonian structure, which takes only the energy into account and not the vortex-related conservation laws.
The use of the Nambu representation to form a Lie algebra for vortex dynamics that is directly based on the vorticity equation can be seen as an advantage. Moreover, the conservation of mass is implicitly included in the Nambu formulation, whereas, concerning the Euler equations of motion and the corresponding Hamiltonian view, the incompressibility condition, i.e. the conservation of mass, is expressed by an additional equation. 
In other words: using the vortex equation, we are already on the appropriate hierarchical level to build a Lie algebra that can be used for the derivation of shortest paths of vortices.
%The steps, how we will proceed to obtain the equations for vortex geodesics for two- and three-dimensional vortex dynamics are summarized in fig. \ref{fig:Subriemann:sketch}. 

%\begin{figure}[t!]
%\begin{framed}
%%\maketitle 
%\begin{center}
%%\begin{LARGE}
%Group representation \\
%$\Downarrow  $ \\
%A group with a differential structure (manifold/tangent space)\\
%\vspace{.3cm}
%\frame{
%Lie Group 
%}\\
%$\Downarrow $  \\
%Tangent space at the identity \\
%\vspace{.3cm}
%\frame{
%Lie Algebra }\\ 
%$\Downarrow $  \\
%Extension of the space \\ %, adding a z-coordinate\\
%%$\Downarrow $  \\
%Adding constraints \\
%$\Downarrow $  \\
%Motion on horizontal planes \\
%%Parallel transport/Covariant derivation $= 0$ \\
%$\Downarrow $  \\
%\vspace{.3cm}
%\frame{
%Geodesics as constrained motion on horizontal planes %(or extremal curves) in the extended space
%}\\
%$\Downarrow $  \\
%Projection on phase space\\
%$\Downarrow $  \\
%\vspace{.3cm}
%\frame{
%Vortex geodesics
%}
%%\end{LARGE}
%\end{center}
%\end{framed}
%\caption{\small{It is sketched how vortex geodesics can be derived via sub-Riemannian geometry.}}
%\label{fig:Subriemann:sketch}
%\end{figure}

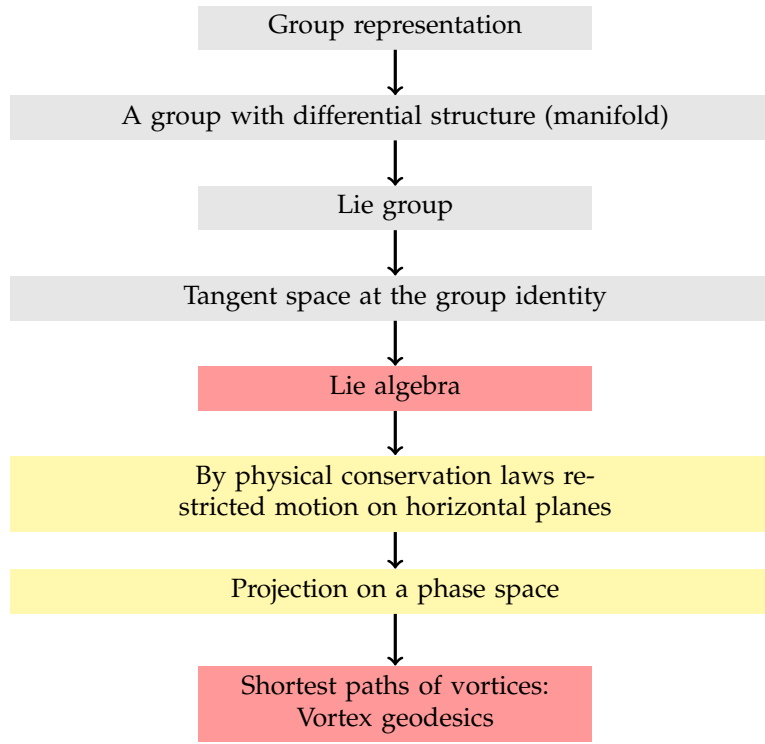
\begin{figure}[t!]
\begin{center}
\begin{tikzpicture}
\node[text width=5cm,align=center] (A) at (0,12) [fill=gray!20]{Group representation};
\node[text width=10cm,align=center] (B) at (0,10.8) [fill=gray!20] {A group with differential structure (manifold)};
\node[text width=5cm,align=center] (C) at (0,9.6) [fill=gray!20]{Lie group};
\node[text width=10cm,align=center] (D) at (0,8.4) [fill=gray!20]{Tangent space at the group identity};
\node[text width=5cm,align=center] (E) at (0,7.2) [fill=red!40]{Lie algebra};
\node[text width=10cm,align=center] (F) at (0,5.8) [fill=yellow!40]{By physical conservation laws restricted motion on horizontal planes};
\node[text width=10cm,align=center] (G) at (0,4.5) [fill=yellow!40]{Projection on a phase space};
\node[text width=5cm,align=center] (H) at (0,3) [fill=red!40]{Shortest paths of vortices: \\ Vortex geodesics};
%\draw[->, blue!50, very thick] (A) to  node[right] {$\nabla \times $} (B);
\draw[->, very thick] (A) to (B);   
\draw[->, very thick] (B) to  (C);  
\draw[->, very thick] (C) to  (D);  
\draw[->, very thick] (D) to  (E);  
\draw[->, very thick] (E) to  (F);  
\draw[->, very thick] (F) to  (G);  
\draw[->, very thick] (G) to (H);  
%\draw (-6.5,9) -- (6.5,9) -- (6.5,1.5) -- (-6.5,1.5) -- (-6.5,9) ;
%\draw (-6.5,1.4) -- (6.5,1.4) -- (6.5,-6) -- (-6.5,-6) -- (-6.5,1.4) ;
\end{tikzpicture}
\caption{\small{How to derive vortex geodesics via sub-Riemannian geometry.}}
\label{fig:Subriemann:sketch}
\end{center}
\end{figure}

%Starting with Lie group, which is a group with an underlying manifold, 
%we can consider a Lie algebra, which can be regarded as the tangent space at the group identity.
%The tangent space, denoted by $TM$, is the space that 
%contains all velocity vectors of all possible curves lying on the manifold $M$.
%Further, we denote with $T_{\mathbf{x}}M$ the tangent space of $M$ in a point $\mathbf{x}\in M$. 
%As discussed in the previous paragraph, regarding the sub-Riemannian structure, standing  on the tangent space gives us the direction where to move, but we have some restriction, where we are allowed to move.  
%Now, we have to split the tangent space of the tangent space (short: $TTM$) into vertical and horizontal subboundles and use the isomorphism $TTM \cong TM$, see \eqref{eq:DirectSum} and \eqref{eq:isomTM-TTM}, because we are only allowed to walk on horizontal planes. From the fluid dynamical perspective, it is natural that vortices do not move on straight lines, they are restricted by the rotational part of the motion leading to constraints with respect to the vortex-related conserved quantities. 

 %For more general explorations and applications of sub-Riemannian geometry see e.g. \citet{Myasnichenko2002}, \citet{Perez2006}, \citet{Montgomery2006}, \citet{Brockett1982}, \citet{Strichartz1986} or \citet{Monroy1999}

\begin{figure}[t]
\centering{
\includegraphics[scale=.6]{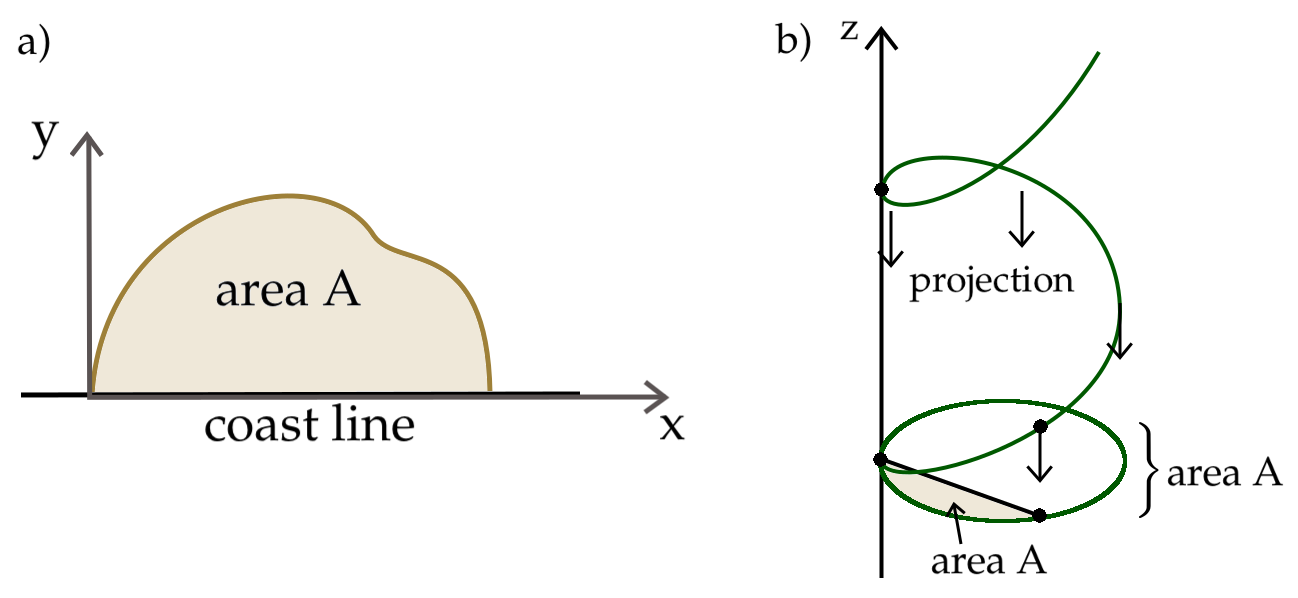} %\includegraphics[scale=.8]{2Wirbel_2D.eps}
\caption{\small{a) Dido's problem was to enclose a maximal area with a leather string of fixed length. b) To solve her problem, she had to extend the space by a $z$-coordinate, and the projection of the 3D curves leads to a solution. Based on \citet{Montgomery2006}.}}
\label{fig:DidosProb}
}
\end{figure}

%We will identify $p$ with the pair $(\mathbf{h},\mathbf{H})$ where as we will show in the following, $\mathbf{H}$ denotes the antisymmetric rotational tensor and $\mathbf{h}$ the vector containing the linear momenta. 
%We see that the left and right hand side of equation \eqref{eq:dpdt} contains $p$, i.e. $(\mathbf{h},\mathbf{H})$  such that solving the exponential function leads to a solution. 
%\color{red} Mehr Quellen! \color{black}

A classical example of sub-Riemannian geometry is Dido's problem \citep[see, e.g.][]{Montgomery2006}. It goes back to the time of the beginning of Rome.
Queen Dido had to flee, arriving at Africa she was allowed to get as much land as she could enclose with a leather string of fixed lengths. She got a piece of land at he coast. Approximating the coast line by a straight line, she had to tackle the question: What is the shape of the curve that encloses a maximal area? This problem is called Dido's problem, illustrated in fig. \ref{fig:DidosProb}, and the solution -- a half circle -- can be derived by the use of sub-Riemannnian geometry. 
 
To find shortest paths on a land surface with the constrained of the enclosure of a maximal area seems like a two-dimensional problem. Consider the coordinates $x,y$ and the differential 1-form (see def. \ref{def:1-form})
\begin{equation}
\omega = \frac{1}{2} (xdy-ydx), 
\end{equation} 
which satisfies $d\omega = dy \wedge dy$. Denote with $A$ the area enclosed by a planar curve $c$ (the leather string of fixed lengths) and a straight line segment (the coast line), where the straight line segment and the curve intersect at the origin. The area $A$ is given by: 
\begin{equation}
A(c) = \int_c \ \omega
\end{equation}
the curve $c$ is a function $c = (x(t),y(t))$ with length 
\begin{equation}
l(c) = \int_c \ ds,
\label{eqLlvonc}
\end{equation}
where $ds = \sqrt{dx^2+dy^2} = \Vert \dot{c} \Vert dt$.

To solve Dido's problem we construct a three-dimensional geometry and add a third direction $z$ such that we can find three dimensional curves of the form $(x(t),y(t),z(t))$. Therefore, we consider the 1-form
%We recall that we search for the admissible paths given by admissible velocities at a point, say at $(x,y,z)$. The subspace of admissible velocities is the kernel of the 1-form $\omega$ such that we can say that a curve 
%$t \mapsto (x(t),y(t),z(t))$ is an admissible path if and only if 
\begin{equation}
\dot{z} =  \frac{1}{2} (-y(t) \dot{x}(t)+x(t) \dot{y}(t)).
\end{equation}
such that the single planar curve $c(t) = (x(t),y(t))$ is linked to a family of 
three-dimensional curves that we denote with $\gamma(t) = (x(t),y(t),z(t))$. 
We define the length of this paths $\gamma $ to be equal to the Euclidean length $l$ of the two-dimensional curve \eqref{eqLlvonc}.
Such three-dimensional paths $\gamma$ are called horizontal path.
We will give a precise mathematical definition later in this work.
Then, each planar curve $(x(t),y(t))$ has a lift to  $(x(t),y(t),z(t))$ in $\mathbb{R}^3$, where $z(t)$ is given by the integral:
\begin{equation}
{z}(t) = z(t)-z(0) =  \frac{1}{2} \int_0^t (x(t) d{y}(t)-y(t) d{x}(t)).
\end{equation}
Or in other words, $c(t)$ is the projection of $\gamma(t)$ to the plane. 
We apply Stoke's theorem, assume the following initial conditions $z(0)=0$, $x(0) = y(0) = 0$ such that $c$ joins the origin and it also joins $(x_1,y_1)$.
Then, for the endpoints of the 3D curve $\gamma$ we obtain ($x_1,y_1,A(c)$.)
Thus, this example yields the solution of Dido's problem: Finding a shortest curve between two points, where the curve together with the straight line between the points enclose a certain -- in this case a maximal -- area \citep[see, e.g.][]{Montgomery2006}.
The algebra is hidden in the part, where we considered the three-dimensional curves as horizontal paths. These paths $\gamma$ are on tangent spaces, and Lie algebras can be regarded as tangent spaces at the group identity. Our motivation of the first example was to illustrate the applicability of sub-Riemannian geometry to a simple problem.

 The vortex geodesics for two-and three-dimensional vortex flows will be derived after the generalized algorithm of  \citet{Perez2006}. The steps for the derivations regarding sub-Riemannian geometry are summarized in fig. \ref{fig:Subriemann:sketch}.
In section \ref{sec:vh23}, we will begin with a short summary of the Nambu formulation of two- and three dimensional incompressible, inviscid fluid flows resulting in the Vortex-Heisenberg algebras vh(2) and vh(3). Followed by an introduction to the basic differential geometric definitions in section \ref{sec:basics}. The basics are needed for the general algorithm to find sub-Riemannian geodesics for a 2-step nilpotent Lie algebra that we will introduce in section \ref{sec:step2allg}. Thereby, we will consider a tangent boundle that is spanned by vector fields. 
These vector fields are associated with Hamiltonian functions that generate the Lie algebra \footnote{
We remark that the Hamiltonian here is not related to the energy. Within this work, we are looking for geodesics and therefore, we apply optimal control theory where the notion Hamiltonian is commonly used for the function which solution provides extremal trajectories, especially under constraints. The Russian mathematician Lev Pontryagin introduced this terminology in 1965.}. 
In section \ref{subsec:2-3subr} We will regard the Vortex-Heisenberg Lie algebra vh(2) and apply the algorithm to the discretized conservation laws, i.e. the point vortex model. 
For each Lie algebra we will obtain a pair $(\mathbf{h},\underline{\mathbf{H}})$ that is composed of a vector $\mathbf{h}$ and a matrix $ \mathbf{H}$. For vh(2) we will obtain a vector $\mathbf{h} \in \mathbb{R}^2$ and 
$\mathbf{H} \in$Mat$(2\times 2)$.
In section \ref{sec:3-6}, we outline the results for the discretized conservation laws of vh(3). 
In this case, we obtain a  vector $\mathbf{h} \in \mathbb{R}^3$ and 
$\mathbf{H} \in$Mat$(3\times 3)$. 
Determining the corresponding hauptspace leads to the phase space of the geodesics. %We will use his maximum (or minimum) principle to find vortex geodesics. 
Finally, in section \ref{sec:summary} the results are summarized. 

\section{The Vortex-Heisenberg algebras vh(2) and vh(3)}
\label{sec:vh23}

In order to apply sub-Riemannian geometry to incompressible vortex dynamics, we recall the
definitions of the Vortex-Heisenerg algebra in two-and three spatial dimensions. Both algebras
result from the Nambu representation of the incompressible vorticity equation \citep{Nevir1993}. A summary of the Nambu-brackets generating the Vortex-Heisenberg algebra
is also given in \citet{muller2021algebra}.

\subsection{The Heisenberg-Vortex Lie algebra vh(2)}

In a first step, the conserved quantities for two-dimensional incompressible, inviscid fluids are formulated as functionals of the vorticity with respect to the infinitesimal area element $df$ and the material surface $F$: 
\begin{eqnarray} \label{eq:Energy2D}
{\cal H}[\zeta]    & = & -\frac{1}{2} \int\limits_F df \psi\zeta  
\hspace{.5cm} \text{(kinetic energy)} \\  \label{eq:energy3D}
{\cal P}_x[\zeta]  & = & +\int\limits_F df y \zeta 
\hspace{1cm} \text{(linear momentum component in $x$-direction)}\\
{\cal P}_y[\zeta]  & = & -\int\limits_F df x \zeta  
\hspace{1cm} \text{(linear momentum component in $y$-direction)}\\
{\cal L}_z[\zeta]  & = & -\frac{1}{2}\int\limits_F df r^2 \zeta 
\hspace{.6cm} \text{(components of the angular momentum)}\\
{\cal Z}[\zeta]    & = & \int\limits_F df \zeta 
\hspace{1.7cm} \text{(circulation)}\\
{\cal E}[\zeta]   & = & \frac{1}{2}\int\limits_F df \zeta^2 
\hspace{1.3cm} \text{(enstrophy)}\label{eq:Enstrophy2D}
\end{eqnarray}
Then, the Nambu bracket for two-dimensional incompressible fluids is calculated with respect to the enstrophy ${\cal E}$. The enstrophy is a positive definite constitutive vortex conservation quantity and can be regarded as an integral measure of the vortical degree of a 2D flow. 
Considering two arbitrary functions ${\cal F}$ and ${\cal G}$ the Nambu bracket is defined as as: 
\begin{equation} \label{eq:Nambu_bracket2D}
\{ \mathcal{F},\mathcal{E},\mathcal{G} \}
=-\int \limits_V \ df \frac{\delta \mathcal{F}}{\delta \zeta}
J \left( \frac{\delta \mathcal{E}}{\delta \zeta },\frac{\delta \mathcal{G}}{\delta \zeta} \right), 
\end{equation}
where $J(.,.)$ defines the classical Jacobi-determinant \citep{Nevir1993}. 

While the temporal evolution can be expressed via the enstrophy and the kinetic energy in the Nambu bracket, calculating the Nambu bracket with respect to the enstrophy identifies spatial changes of the structure. 
Following \citet{Nevir1998}, the brackets of the above conserved quantities results in:
\begin{equation} 
\begin{split}
\{{\cal P}_x,{\cal E},{\cal P}_y\}=  {\cal Z} ,\hspace{1.9cm} \{{\cal L}_z,{\cal E},{\cal P}_x\}&=+ {\cal P}_y ,\hspace{2cm} \{{\cal P}_x,{\cal E},{\cal H}\}= 0  \\ 
\{{\cal P}_x,{\cal E},{\cal Z}\}= 0 ,\hspace{2cm} \{{\cal L}_z,{\cal E},{\cal P}_y\}&=- {\cal P}_x  ,\hspace{2cm} \{{\cal P}_y,{\cal E},{\cal H}\}= 0  \\
\{{\cal P}_y,{\cal E},{\cal Z}\}= 0,\hspace{2.1cm} \{{\cal L}_z,{\cal E},{\cal Z}\}&= 0,\hspace{2.6cm} \{{\cal L}_z,{\cal E},{\cal H}\}= 0 \, .   \\ 
\end{split} \label{eq:Bracket_relations_2D}
\end{equation}
The block on the right hand side will be considered for the application of sub-Riemannian geometry.
In general, a $(n, n(n+1)/2)$- sub-Riemannian geometry can be applied to a $(n(n+1)/2)$-dimensional Lie algebra with a rank $n$ distribution.
A distribution $\Delta$ of rank $n$ is  a subspace of the tangent space of a manifold, where there exist $n$ linearly independent vector fields that form a basis for the distribution $\Delta$.
For vh(2), we have seen that the corresponding Nambu brackets of the components of the momentum and the circulation result in the circulation.

Thus, the rank of the Lie algebra is three (the two momentum components and the circulation) with rank of distribution $2$, which is the number of elements (number of components of the momentum).
%$(n,n(n+1)/2):. n(n+1)/2)$: Dimension der algebra with rank n distribution, a distribution $\Delta$ is  a subspace of the tangent space, where there exist n linnearly indepedent vector fields that form a basis for the subspace $\Delta$, see def. ? 
We will consider the discretized conservation laws and show, how the from the Vortex-Heisenberg algebra resulting two-dimensional sub-Riemannian vortex geodesics can be related to a special point vortex constellation -- a relative equilibrium.

\subsection{The Heisenberg-Vortex Lie algebra vh(3)}

Analogously to the the previous subsection, we first recall the volume integrals that constrain the fluid dynamical system.
We denote ${\bf A}$ as the vector potential $\mathbf{v} = - \nabla \times {\bf A}$, which is a solution of $\nabla \cdot \mathbf{v} = 0$. 
The latter equation is a central condition for the Nambu representation that contains the conservation of mass in the algebraic framework. 
The kinetic energy, the linear momentum
%%% HERE
$\boldsymbol{\mathscr{P}} = ({\cal P} _x,{\cal P} _y,{\cal P} _z)$ and the angular momentum $\boldsymbol{\mathscr{L}} = ({\cal L} _x,{\cal L} _y,{\cal L} _z)$ are classical conservation laws of dynamical systems. 
These conserved quantities are formulated as functionals of the vorticity vector $\boldsymbol{\xi}$ with respect to the infinitesimal volume element $d\tau$ and volume $V$:
\begin{eqnarray} \label{eq:3D_conservationLaws}
{\cal H}[\boldsymbol{\xi}]& = &-\frac{1}{2}\int\limits_V d \tau\; {\bf A}\cdot\boldsymbol{\xi} 
\hspace{1.8cm} \text{(kinetic energy)} \\
\boldsymbol{\mathscr{P}}[\boldsymbol{\xi}]& = &\frac{1}{2}\int\limits_V d \tau\; ({\bf r}\times \boldsymbol{\xi}) 
\hspace{2cm} \text{(linear momentum)} \\
\boldsymbol{\mathscr{L}}[\boldsymbol{\xi}]& = & -\frac{1}{2} \int \limits_V d \tau\; r^2 \boldsymbol{\xi} 
\hspace{2.1cm} \text{(angular momentum)} \\
\boldsymbol{\mathscr{Z}}[\boldsymbol{\xi}]& = & \int\limits_V d \tau\; \boldsymbol{\xi}
\hspace{3.2cm} \text{(total flux of vorticity)}\\
h [\boldsymbol{\xi}]& = & \frac{1}{2}\int\limits_V d \tau\; { \bf v}\cdot \boldsymbol{\xi}
\hspace{2.5cm} \text{(helicity)}
\end{eqnarray}
\citep[See also][p. 25]{Majda2002}. In three spatial dimensions, the trilinear Nambu bracket can be defined as:
\begin{equation}\label{eq:vorticityeqwK}
\frac{\partial \mathcal{F}}{\partial t} =
\{ \mathcal{F}, h, \mathcal{H}\} :=
 \int_V d \tau \frac{\delta \mathcal{F}}{\delta  \boldsymbol{\xi}} \cdot \frac{\partial \boldsymbol{\xi}}{\partial t}
= \int_V d \tau \frac{\delta \mathcal{F}}{\delta  \boldsymbol{\xi}} \cdot  \mathbf{K}
\left( \frac{\delta h}{\delta  \boldsymbol{\xi}},\frac{\delta \mathcal{H}}{\delta  \boldsymbol{\xi}} \right),
\end{equation}
respectively, 
\begin{equation} 
\{{\cal F}, h,{\cal G}\}=-\int_{V} d\tau\left[\left(\nabla\times\frac{\delta {\cal F}}{\delta \boldsymbol{\xi}}\right)\times\left(\nabla \times 
\frac{\delta h}{\delta \mbox{\boldmath $\xi$}} \right) \cdot\left(\nabla\times
\frac{\delta {\cal G}}{\delta \boldsymbol \xi}\right)\right] 
\end{equation}
The Nambu bracket is antisymmetric in all arguments. This property follows from the triple and cross products. The Nambu bracket is also multilinear. And keeping the helicity $h$ in the middle argument fixed, the Nambu bracket can be reduced to a Poisson bracket that satisfies the Jacobi identity, see \citet{Nevir1993} and \citet{Nevir1998}.

%%%%%%%%%%%%%%%%%%%%%%%%%%%%%%%%%%%%%%%%%%%%%

Determining the Nambu brackets of the conserved quantities for incompressible flows leads to the following bracket relations:
\color{black}
\begin{equation}\label{eq:Bracket_relations1}
\begin{split}
\{ {\cal P}_{\alpha},h,{\cal P}_{\beta}\} &=\varepsilon_{\alpha\beta\gamma} {\cal Z}_{\gamma}, 
\hspace{2.9cm} \{{\cal L}_{\alpha},h,{\cal L}_{\beta}\} =\varepsilon_{\alpha\beta\gamma} {\cal L}_{\gamma} \\
\{{\cal P}_{\alpha},h,{\cal Z}_{\beta}\}&= 0 , 
\hspace{3.85cm} \{{\cal L}_{\alpha},h,{\cal P}_{\beta}\}=\varepsilon_{\alpha\beta\gamma} {\cal P}_{\gamma} \\
\{ {\cal Z}_{\alpha},h,{\cal Z}_{\beta}\}&= 0 \\
& \\
\{{\cal P}_{\alpha},h,{\cal H }\} &= 0 , \hspace{3.85cm}  \{{\cal L}_{\alpha},h,{\cal Z}_{\beta}\} = 0  \\
 \{{\cal Z}_{\alpha},h,{\cal H}\} &= 0 , \hspace{3.85cm}  \{{\cal L}_{\alpha},h,{\cal H}\}= 0  \\
\end{split}
\end{equation}
with the Levi-Civita symbol $\varepsilon_{\alpha\beta\gamma} $ and $\alpha,\beta,\gamma \in \{1,2,3\}$ \citep[see][]{Nevir1998}.

Within this work, we will concentrate on the upper block on the left hand side. We note that
the bracket of the components of the momentum do not commute, which is contrary to mass point dynamics. 
The dimension of the Vortex-Heisenberg Lie algebra is $6$ 
(6 elements: the components of the linear momentum and three components of the total flux of vorticity) with rank $3$ distribution. The Heisenberg-Vortex Lie algebra. Analogously to the application of vh(2) we will consider the discretized conservation laws.
Therefore, vh(3) yields an example of a (3,6)-sub-Riemannian geometry. 
But this application should be seen as an outlook, which will be more examined in future studies. In two dimensions, we can compare the (2,3)-sub-Riemannian geodesics with a discrete point vortex model. Such an idealized discrete model does not exist for three dimensions.

%Consider the upper block on the left hand side, where the Nambu bracket of the components of the total flux of vorticity ${\cal Z}_{\alpha}$ and the momentum ${\cal P}_{\alpha}$, $\alpha = 1,2,3$, with respect to the helicity \eqref{eq:3D_conservationLaws} is calculated.
%Comparing this bracket to the Poisson bracket for mass point dynamics, we notice that the brackets concerning the components of the linear momentum of mass points commute generating an Abelian algebra. In contrast, regarding fluid dynamics, the Nambu bracket of the components of the momentum is the total flux of vorticity generating a nilpotent Lie algebra, as it will discuss in section \ref{sec:algebra3D}. Therefore, this Nambu bracket is a central extension of the translation algebra. 

\section{Basics of differential geometry}
\label{sec:basics}

%We can either consider the whole point vortex system as one system, i.e. $\Gamma = \sum^N \Gamma$ leading the the following matrix:    
%\begin{equation}
%\mathbf{H} = \begin{pmatrix} 0 & \Gamma \\ -\Gamma & 0 \end{pmatrix},
%\end{equation}
%alternatively, we consider $N$ systems whereby each point vortex $i$ with circulation $\Gamma_i$ lives in its own eigenspace:
%\begin{equation}
%\mathbf{H}_i = \begin{pmatrix} 0 & \Gamma_i \\ -\Gamma_i & 0 \end{pmatrix}
%\end{equation}  
%With this matrices of rotations we can calculate the Eigenspace and then the horizontal extremal curve on TM (TTM).
%As last step we project the horizontal extremal curve into the phase space and the image is our sought-after vortex trajectory! 
%
%
We shortly summarize the basics of differential geometry.
For more detailed proofs and explanations see e.g. \citet{Kuhnel1999}, \citet{Berger2012} or \citet{DoCarmo2017}.

%\begin{itemize}
% \item $\mathbb{R}^n \times so(n)$ system
% \item $(\dot{x},\dot{z})$, $\dot{x}$: vector, $\dot{z}$: skew-symmetric matrix
% \item $H_{n+1}$: Lie group with underlying manifold $\mathbb{R}^n \times so(n)$
% \item $H_3$: 3D Heisenberggroup $\mathbb{R}^2 \times so(2)$
% \item $\dot{x}=u$, $\dot{z} = x \wedge u$
% \item \textit{We set the problem on a nilpotent Lie group.
% It is known that if the group is simply connected, then the exponential
% napping establishes an analytic diffeomorphism from the algebra onto the group}
% \item \color{red} Goal: Finding the minimum x --- the geodesic arcs of the system \color{black}
% \end{itemize}

\begin{definition}\textbf{Tangent space}\\
The tangent space, denoted by $TM$, is the space that 
contains all velocity vectors of all possible curves lying on the manifold $M$.
With $T_{\mathbf{x}}M$ we denote the tangent space of $M$ in a point $\mathbf{x}\in M$. %\color{blue} p als Ort bezeichnen? \color{black}
\end{definition}

\begin{definition}\textbf{Linear form and dual space}\\
Let $V$ be a vector space over a field $K$. Then, a linear functional or linear form 
(which is also called one-form or covector) is a map from $V$ to $K$ that is linear 
and satisfies the following conditions:
\begin{equation}
\begin{split}
\varphi(\mathbf{v}+\mathbf{w}) & = \varphi(\mathbf{v})+\varphi(\mathbf{w})\hspace{1cm} \text{for all }\mathbf{v},
\mathbf{w}\in V \\
\varphi(a\mathbf{v}) &= a \varphi(\mathbf{v}) \hspace{2.25cm} \text{for all } \mathbf{v}\in V, a\in K.
\end{split}
\end{equation}
All linear functionals from $V$ to $K$ form a vector space Hom$(V,K)$ over the field $K$. This space
\begin{equation}
V^{\ast} := \text{Hom}(V,K) = \{ \phi:V \longrightarrow K \vert \ \phi \text{ linear } \} 
\end{equation} 
 is called the dual space of $V$.
\end{definition}
We notice that dealing with Lie algebras means dealing with linear spaces. Therefore, to find vortex geodesics based on the Vortex-Heisenberg algebra, we need to introduce the following definitions.

\begin{definition}\textbf{Pfaffsche Form}\label{def:1-form}\\
A Pfaffsche Form $\omega$ on $M$ (also called 1-Form or differential form) assigns a linear form 
$ \omega_p\colon\mathrm T_{\mathbf{x}}M\to\mathbb R $
to each point $\mathbf{x}\in M$.

Such linear forms are called cotangent vectors which are elements of the so-called dual space (or cotangent space) $\mathrm T^{\ast}_{\mathbf{x}}M$  with respect to the tangent space $\mathrm T_pM$. 
Therefore, a Pfaffsche Form $\omega$ is a map
\begin{equation}
    \omega\colon M\to\bigcup_{\mathbf{x}\in M}\mathrm T^*_{\mathbf{x}}M,\quad \mathbf{x}\mapsto\omega_{\mathbf{x}}\in\mathrm T^*_{\mathbf{x}}M. 
\end{equation} 
\end{definition}

\begin{example}
A 1-form $\phi$ on $\mathbb{R}^3$ is a function on the set of all tangent vectors to $\mathbb{R}^3$ such that $\phi_p$ is is linear using the notation $\phi_p := \phi(p)$.
This means, for $\alpha,\beta \in \mathbb{R}$ and tangent vectors $\mathbf{v},\mathbf{w} \in \mathbb{R}^3$ it is:
\begin{equation}
\phi_p (\alpha \mathbf{v}+b\mathbf{w}) = \alpha \phi_p(\mathbf{v})+ \beta \phi_p(\mathbf{w}) 
\end{equation}
We follow for example \citet{ONeill2006} and emphasize that for every tangent vector $\mathbf{v}$, $\phi_p$ maps to a real number. 
Moreover, for each point $\mathbf{x} \in \mathbb{R}^3$, the resulting function $\phi_{\mathbf{x}}: T_{\mathbf{x}}\mathbb{R}^3 \rightarrow \mathbb{R}$ is linear. 
That means that at each point $\mathbf{x}$, $\phi_{\mathbf{x}}$ is an element of the dual space of $T_{\mathbf{x}}\mathbb{R}^3$
\end{example}
%Let us consider the euclidean plane $\mathbb{R}^2$ and shift a vector on a straight line without rotating it. 
%Two vectors are called parallel, if one vector exactly matches another vector after shifting, 
%Moreover, the straight line is the geodesic, i.e. the minimum distance 
%between these two vectors in the euclidean space. 

\begin{definition}\textbf{Parallel vector field}\\
A vector field $\mathbf{V}(t)$ along a regular parametrized curve $\gamma$ of constant length 
is called parallel if the derivative $\mathbf{V}'(t)$ is normal to the tangent plane 
$T_{\mathbf{x}}M$ 
at each point $\mathbf{x} = \gamma(t)$ of the curve. 
\end{definition}
If for all $t$ the parallel field $\mathbf{V}'(t)$ is normal to the tangent plane,
the length of the vectors $\mathbf{V}(t)$ is constant, which follows immediately from:
\begin{equation}
\frac{d}{dt} \vert \mathbf{V} \vert ^2 = \frac{d}{dt} (\mathbf{V} \cdot \mathbf{V}) 
= 2\mathbf{V}' \cdot \mathbf{V} = 0.
\end{equation}

\begin{definition}\textbf{Geodesics}\\
A unit speed curve $\gamma$ on a surface M is a geodesic if and only if its tangent vectors 
$\gamma '(t)$ form a parallel field. 
\end{definition}

\begin{example} Let us consider the great circle on a 2-sphere. If we move along the geodesic, i.e. along the great circle parametrized as regular curve, we recognize that the angle of the tangent vector is constant. Therefore, the tangent vectors of the great circle form a parallel field. Thus, the great circle is a geodesic curve on the 2-sphere.
\end{example}

The concept of the covariant derivative can be used to determine directional derivatives of vector fields, i.e. the infinitesimal transport of a vector field in a given direction.

\begin{definition}\textbf{Covariant derivative and parallel transport} \\
Let $\mathbf{V}$ be a smooth vector field along a curve $c:I \rightarrow M$ on a manifold. 
Then the covariant derivative of $\mathbf{V}$ along $c$ at the point $p=c(t)$
is given by 
\begin{equation}
\frac{D_c \mathbf{V}}{dt} (p) = \frac{D_c \mathbf{V}}{dt} (c(t)) = \lim_{s \rightarrow t} \frac{P_{c(t),c(s)} \mathbf{V}_{c(s)} - \mathbf{V}_{c(t)}  }{s-t} \ \in T_pM
\end{equation}
with parallel transport $P_{c(t),c(s)} $ of the tangent vector $\mathbf{V}_{c(s)}$ 
to the tangent vector at the point $c(t)$. 
\end{definition}

We can picture the covariant derivative $\frac{D_c \mathbf{V}}{dt} (c(t)) $ 
as the projection of $\frac{d \mathbf{V}}{dt} $ into the tangent plane to the surface. 
Physically, the covariant derivative $\frac{D_c \mathbf{V} }{dt} (c(t)) $ of a particle
trajectory $c(t)$ along the surface with velocity field 
%\color{blue}$\mathbf{V} = \frac{d c}{dt}$ (nochmal nachschlagen, ob V so definiert sein kann!) \color{black}
represents the acceleration component of the particle along the surface.

Let now $TM$ denote the tangent bundle,  $T_{\mathbf{x}}M$ the tangent space at the point $\mathbf{x}$ and $TTM$ ($T_{\mathbf{x}}TM$) the tangent space of the tangent space (at the point $\mathbf{x}$).
Consider the maps $\theta$ and $d\theta$: 
\begin{equation}
\theta:TM \rightarrow T \quad \text{and} \quad d\theta : TTM \rightarrow TM 
\label{eq:isomTM-TTM}
\end{equation}
Using the notation $d\theta (\mathbf{x}):= d_{\mathbf{x}} \theta$ the kernel and the image of $d\theta$ of $d_{\mathbf{x}} \theta$ are given by:
\begin{equation}
\begin{split}
{\rm Ker}(d_{\mathbf{x}} \theta) & = \{ \mathbf{N} \in T_{\mathbf{x}} TM \ \vert \  d\theta(\mathbf{N}) 
= \mathds{1}_{T_{\theta (\mathbf{x})} }M \in T_{ \theta (\mathbf{x}) } M   \} \\
{\rm Im}(d_{\mathbf{x}}\theta) &= \{ d_{\mathbf{x}} \theta(\mathbf{N}) \ \vert \ \mathbf{N} \in T_{\mathbf{x}} M )  \}
\end{split}
\end{equation}
We remark that $\ker (d_{\mathbf{x}}\theta) \subseteq TT M$ and ${\rm im} (d_{\mathbf{x}}\theta) \subseteq TM$ such that there exists an isomorphisms $\varphi: TTM \rightarrow TM$ such that $TTM \cong TM$, $T_{\mathbf{x}}T M \cong T_{\theta(\mathbf{x})}M$ respectively, as illustrated in the following diagram:

\begin{center}
\begin{tikzpicture}
  \matrix (m) [matrix of math nodes,row sep=3em,column sep=4em,minimum width=2em]
  {
     TTM & TM \\
     TM & TTM \\ };
  \path[-stealth]
    (m-1-1) edge node [left] {$\varphi$} (m-2-1)
            edge node [above] {$d_{\mathbf{x}}\theta$} (m-1-2)
    (m-2-1) edge node [above] {$\varphi ^{-1}$} (m-2-2);
            %node [above] {$\exists$} (m-2-2)
    %(m-1-2) edge node [right] {$\mathcal{B}_T$} (m-2-2);
\end{tikzpicture}
\end{center}

Now, we can identify:
\begin{equation}
\varphi (\ker (d_{\mathbf{x}}\theta)) \cong \ker (d_{\mathbf{x}}\theta)
\quad \text{and} \quad
\varphi^{-1} ({\rm im}(d_{\mathbf{x}}\theta)) \cong {\rm im}(d_{\mathbf{x}}\theta)
\end{equation}
%
%
%both the kernel and the image of $d\theta$  map to the tangent space:
%\begin{equation}
%\begin{split}
%T_{\mathbf{x}}TM &  \xrightarrow{{\rm Ker}(d\theta)} T_{\theta(\mathbf{x})}M\\
%T_{\mathbf{x}}TM &  \xrightarrow{{\rm Im}(d\theta)} T_{\theta(p)}M\end{split}
%\end{equation}
%with dim$(TM) = $dim(ker($TM$))$+$dim(Im($TM$)).

%Now, we take a subboundle, i.e. a subset, of the map TTM $\rightarrow TM$ and 
% set
%\begin{equation}
%V_{\mathbf{x}}:= {\rm Ker}(d_{\mathbf{x}}\theta) = T_{\mathbf{x}}T_{\theta(\mathbf{x})} \subset T_{\mathbf{x}}TM  \quad \text{and } H_{\mathbf{x}} = r_{\mathbf{x}}^{\perp}
%\end{equation}
%where the orthogonal complement $r_{\mathbf{x}}^{\perp}$ is with respect to the usual euclidean inner product. 
and set
\begin{equation}
V_{\mathbf{x}}:= {\ker}(d_{\mathbf{x}}\theta) 
 \quad \text{and}  \quad 
 H_{\mathbf{x}} = V_{\mathbf{x}}^{\perp}.
\end{equation}
Then, we can write:
\begin{equation}
\boxed{T_{\mathbf{x}}TM = H_{\mathbf{x}} \oplus V_{\mathbf{x}} 
}
\label{eq:DirectSum}
\end{equation}
%where $H$ is called \textit{horizontal vector bundle} and $V$ is called \textit{vertical vector bundle}.
%This provides the isomorphsim 
%\begin{equation}
%T_{\mathbf{x}}T_{\theta({\mathbf{x}})}M \cong T_{\theta(\mathbf{x})}M.
%\end{equation}
%with respect to the map $\mathbf{v}: V \rightarrow TM$, respectively:
%$\mathbf{v}_{\mathbf{x}}: V_{\mathbf{x}} \rightarrow T_{\mathbf{x}} M$.

Further, denote $\mathbf{c}\Vert_0^t \in T_{c(t)}M$ the vector that we obtain by shifting the tangent vector $\mathbf{x}$
parallel along $\gamma \vert_{[0,t]}$, where $\gamma$ is a geodesic.
Then we get an isomorphism 
\begin{equation}
h_{\gamma, \mathbf{x}}^{-1}: TM  \rightarrow T_{\mathbf{x}}TM, \quad
\mathbf{v}  \mapsto \frac{\partial}{\partial t} \Big|_{t=0} \left( \mathbf{c}_\mathbf{x}\Vert_0^t \right).
\end{equation}
And the image of the isomorphism $h_{\gamma, \mathbf{x}}^{-1}$ can be identified with $H_{\mathbf{x}}$, i.e. 
\begin{equation}
\text{Im}(h_{\gamma, \mathbf{x}}^{-1}) \cong H_{\mathbf{x}}
\end{equation}
leading to the following definition of $H_{\mathbf{x}}$:
 \begin{equation}
\begin{split}
H_{\mathbf{x}}:= \{ \dot{\mathbf{c}}(0)  \ \vert & \ \mathbf{c}: I \rightarrow \text{TM parallel along a geodesic } \gamma, \\
& \gamma:I \mapsto M \text{ with } \gamma(0)= \theta(\mathbf{x}) \text{ und } \mathbf{X}(0) = \mathbf{x} \} .
\end{split}
\end{equation}
which corroborates the above representation of $T_{\mathbf{x}}TM$ as direct sum of $H_{\mathbf{x}}$ and $V_{\mathbf{x}}$.

%In (\ref{eq:Sp(2n)}), we have defined the symplectic group based on the bilinear form (\ref{eq:SymForm}) and we have discussed the symplectic structure of classical Hamiltonian systems. 
%To extend the symplectic group such that it becomes a Lie group, i.e. a group which is also a manifold, we introduce the terminology of 2-forms on a surface \citep{ONeill2006} leading to the definition of symplectic manifolds.
%
%\begin{definition}
%A 2-form $\eta$ on a surface $M$ is a real-valued function on all ordered pairs of tangent vectors $\mathbf{v},\mathbf{w}$ to M such that
%\begin{enumerate}
%\item $\eta (\mathbf{v},\mathbf{w})$ is linear in $\mathbf{v}$ and $\mathbf{w}$
%\item $\eta (\mathbf{v},\mathbf{w}) = -\eta (\mathbf{w},\mathbf{v}) $
%\end{enumerate}
%\end{definition}
%
%\begin{definition}\textbf{Symplectic manifold}\\
%A symplectic manifold is a smooth manifold $M$ that is equipped with a closed non-degenerate differential 2-form $\eta$ .
%\end{definition}

\section[Sub-Riemannian Geometry]{Sub-Riemannian geometry of a  step-2 nilpotent Lie algebra}
\label{sec:step2allg}

We will use this algebraic analysis to derive vortex geodesics. In general, sub-Riemannian geometry can be used to study constrained physical systems. Regarding hydrodynamical systems, the conservation of the vortex quantities restrict the motion on the tangent space. Furthermore, vh(2) and vh(3) are nilpotent algebras. Both lead to a natural sub-Riemannian structure for vortex dynamics. 
In this section we will summarize, how sub-Riemannian geodesics can be determined for general step-2 nilpotent algebras. In sections \ref{sec:2.3allg} and \ref{sec:3-6} we will apply this proceeding to the two- and three-dimensional Vortex-Heiseberg algebras vh(2) and vh(3).

%After the introduction to the basic definitions in differential geometry, we have the basic ingredients to discuss sub-Riemannian geometry.
Denote $\mathfrak{g}$ a arbitrary step-2 nilpotent Lie-Algebra with respect to the step-2 nilpotent Lie group $G$ and
denote $\{ \mathbf{X}_i, \mathbf{X}_{jk} \ \vert \  i=1,\dots ,n,\ 1 \leq j <k \leq n \}$ the basis of the $n(n+1)/2$-dimensional Lie Algebra $\mathfrak{g}$ with multiplication table:
\begin{equation}
[\mathbf{X}_i,\mathbf{X}_j] = \mathbf{X}_{ij},\ [\mathbf{X}_i,\mathbf{X}_{jk}] = 0,\ [\mathbf{X}_{ij},\mathbf{X}_{kl}] = 0.
\end{equation}
In order to derive the $(n,n(n+1)/2)$-sub-Rimannian geodesics with respect to the Lie algebra, we consider $\mathfrak{g}$ as a family of left-invariant vector fields on the Lie group $G$. 
Then $\Delta = \text{span}(\mathbf{X}_1,\dots , \mathbf{X}_n)$ is a left invariant bracket generating distribution on the Lie group $G$ and of rank $n$.
Further, we assume that the vectors $\mathbf{X}_i(g),~\ i=1,\dots ,n$ are orthogonal such that we can define an inner product on the plane
$\Delta (g) =  \text{span}(\mathbf{X}_1(g),\dots , \mathbf{X}_n(g))$ that varies smoothly with respect to $g$. 

\begin{definition}\textbf{sub-Riemannian distance, sub-Riemannian length} \\
Let $g:[0,T] \longrightarrow G$ be a horizontal curve, then $\dot{g} \in \Delta(g)$ almost everywhere.
Let $\mathbf{x},\mathbf{x}_T \in G$. Then the sub-Riemannian distance is defined as
\begin{equation}
d(\mathbf{x},\mathbf{x}_T) = \inf\{ \ l(g) \ \vert \ g:[0,T] \longrightarrow G \text{ is horizontal, } g(0)=\mathbf{x}, \  g(T) = \mathbf{x}_T\}
\end{equation}
with the sub-Riemann length $l$ of the curve $g$:
\begin{equation}
l(g) = \int_0^T \Vert \dot{g}(t) \Vert dt.
\label{eq:l}
\end{equation}
with respect to the inner product defined by $X_i(g)$ as explained before the definition.
\end{definition}
We are looking for the Sub-Riemannian geodesics on the group $G$, i.e. for the minimization of the functional in the class of horizontal curves. 
%Quelle: http://people.sissa.it/~percacci/lectures/genrel/05-liegroups.pdf
We recall that a Lie group is a group with a manifold structure and a Lie algebra corresponds to the tangent space at the identity group element. 

The following paragraph is summarized after \citet{Percacci2017}.
Let us call the Lie group $G$ and let the multiplication $ \ast : G \times G \rightarrow G $ and the inverse $I : G  \rightarrow  G$ are smooth maps. 
We consider the diffeomorphisms $L_g : G \rightarrow  G$ defined by $L_g(g') = gg' $ satisfying the composition property of the Lie group. 
Then, a vector field $\mathbf{v} \in \mathbf{X}(G)$ is called left-invariant if $T L_g(\mathbf{v} ) =\mathbf{v}$ for all $g \in G$. 
We note that (i) the Lie bracket of two left-invariant vector fields is a left-invariant vector field, and (ii) right-invariant vector fields can be defined analogously. 
As we have discussed before with respect to the Lie bracket, the left-invariant vector fields form a Lie algebra $L(G)$ of the Lie group $G$. 
Denote the identity element of the group with $e$ and with $\bar{\mathbf{v}} \in T_eG$ the vector tangent to the group $G$ at the identity such that can define a unique left-invariant vector field $\mathbf{v} $ that coincides with $ \bar{\mathbf{v}}$  in the identity: 
\begin{equation}
\mathbf{v} (g) = T L_g(\bar{\mathbf{v}} )
\end{equation}
Therefore, there is a one-to-one correspondence between elements of $T_eG$ and left-invariant vector fields such that the dimension of the Lie algebra $L(G)$ is equal to dim($G$). 
We recall that a Lie bracket of two vector fields $\mathbf{X}$ and $\mathbf{Y}$ is left-translation invariant if $\mathbf{X}$ and $\mathbf{Y}$ are left-invariant.  

We denote with $g \mapsto L_g$ the left-action on the group $G$ that defines tangent and cotangent bundle trivializations.  Then we can write:
\begin{equation}
G \times \mathfrak{g}\simeq TG, \qquad G \times \mathfrak{g^{\ast}}\simeq T^{\ast}G,
\end{equation}
with respect to the mappings
\begin{equation}
(\mathbf{g},\mathbf{X}) \mapsto d_e L_g \mathbf{X}, \qquad (\mathbf{g},\mathbf{x})) \mapsto d_e L_{g^-1} (g)^{\ast}\mathbf{x}.
\end{equation}
We remind that the space of left-invariant vector fields can be identified with the tangent space of the group identity such that we can relate this notation to the Lie bracket.
Using the isomorphism $\varphi: TTM \rightarrow TM$ explained in section \ref{sec:basics}, i.e. $TTM \cong TM$, we obtain the double bundle 
\begin{equation}
TT^{\ast}G \simeq (G\times \mathfrak{g}^{\ast} ) \times (\mathfrak{g} \times \mathfrak{g}^{\ast}).
\end{equation}
Now, we can represent any tangent vector as pair $((g,p),(\mathbf{X},\mathbf{Y}^{\ast}))$ and the symplectic form $\omega_{(g,\mathbf{x})}$ is given by:
\begin{equation}
\omega_{(g,\mathbf{x})} ((\mathbf{X}_1,\mathbf{Y}_1^{\ast}),(\mathbf{X}_2,\mathbf{Y}_2^{\ast})) )
 = \mathbf{Y}_1^{\ast} (\mathbf{X}_2) -  \mathbf{Y}_2^{\ast} (\mathbf{X}_1) - \mathbf{x}[\mathbf{X}_1 , \mathbf{X}_2] \, .
\end{equation}
See \citet{Perez2006} for further details. 

Let now $H: T^{\ast}G \longrightarrow \mathbb{R} $ be a Hamiltonian function. As we have noticed earlier in this context the Hamiltonian does not represent the energy.
It is a vector field which solution provides the geodesics. 
The flow of the Hamiltonian vector field ${H}$ is given by the system:
\begin{equation}
\begin{split}
\frac{d\mathbf{g}}{dt} & = (dL_g)(dH) \\
\frac{d\mathbf{x}}{dt} &  = -(\text{ad}^{\ast}dH)(\mathbf{x}),
\label{lgs}
\end{split}
\end{equation}
Let now $ H_i $ be the Hamiltonian function with respect to the vector field $\mathbf{X}_i$ %, i.e. $ H_i  = \mathcal{H}_{X_i}$  
%\color{blue} was bedeutet dieser Zusammenhang? \color{black} 
and denote 
$H_{ij}$ the Hamilton function corresponding to $\mathbf{X}_{ij}$, that means $ H_{ij}  = \mathcal{H}_{\mathbf{X}_{ij}}$.
The Lie Poisson bracket is given by
\begin{equation}
[H_i,H_j] = H_{ij},
\label{eq:bracket}
\end{equation}
where $H_{ij}$ are central elements of the Lie algebra $T^{\ast} \mathfrak{g}$ generated by the bracket relation \eqref{eq:bracket}.
These Hamiltonians depend only on the second variable $\mathbf{x}$, because of the left invariance of the vector fields. 

Further, denote $\{\mathbf{X}_i^{\ast}, \mathbf{X}_j^{\ast} \}$ the dual basis of the basis $\{\mathbf{X}_i, \mathbf{X}_j \} $.
Applying (\ref{lgs}) the dual variable $\mathbf{x}$ can be identified with the vector
\begin{equation} %\begin{split}
\mathbf{x} \simeq 
  \sum_{i=1}^n \mathbf{x}(\mathbf{X}_i) \mathbf{X}_i^{\ast}
  + \sum_{i<j}^n \mathbf{x}(\mathbf{X}_{ij}) \mathbf{X}_{ij}^{\ast} 
= \sum_{i=1}^n H_i \mathbf{X}_i^{\ast}+ \sum_{i<j} ^n 
H_{ij}  \mathbf{X}_{ij}^{\ast}.
\end{equation}
Moreover, because of the commutator \eqref{eq:bracket}, $\mathbf{x}$ can be identified with the pair 
\begin{equation}
\mathbf{x} \simeq ( \mathbf{h}, \underline{\mathbf{H}})\in \mathbb{R}^n \times \mathfrak{so}(n)
\label{eq:pair}
\end{equation}
with 
\begin{equation}
\mathbf{h} =  \begin{pmatrix} H_1\\ H_2 \\ \vdots \\ H_n
\end{pmatrix}
\quad \text{and} \quad
 \underline{\mathbf{H}}
= \begin{pmatrix} 
H_{11}	& H_{12}	& \dots	 & H_{1n}      \\
H_{21}	& H_{22} 	& \dots  & H_{2n} 	  \\
\vdots	&    	& \ddots & \vdots \\
H_{n1}	& \dots & \dots	 & H_{nn}
\end{pmatrix}
\end{equation}

Using the above identification and \eqref{lgs} the geodesics are completely described by the second equation of the following system \citep{Perez2006}:
\begin{equation*}
\frac{d \mathbf{g}}{dt} = H_1 \mathbf{X}_1(\mathbf{g}) + \cdots + H_n \mathbf{X}_n(\mathbf{g}) 
\end{equation*}
\begin{equation}
\boxed{
 \frac{d\mathbf{x}}{dt}  = (\mathbf{\dot{h}}, \dot{ \underline{\mathbf{H}}}) = ( \underline{\mathbf{H}}\mathbf{h}, \mathbf{0})
} \, .
\label{eq:normed_curve}
\end{equation}
Because $\mathbf{x}$ can be identified with the pair $( \mathbf{h},  \underline{\mathbf{H}})$, we can take the exponential ansatz to solve \eqref{eq:normed_curve}.
Taking the initial condition $\mathbf{h}_0 = \mathbf{h}(0)$ the integral curves are given by
\begin{equation}
t \mapsto (\mathbf{h}(t), \underline{\mathbf{H}}(t)) = \exp(t \underline{\mathbf{H}}\mathbf{h}_0,  \underline{\mathbf{H}}(0)).
\end{equation}
The calculation of the exponential function leads to the geodesic equation. 
To solve the exponential function we apply Lagrange's and Sylvester's formula that represents an analytic function $f(A)$ of a  diagonalizable $n \times n$-matrix $ \underline{\mathbf{A}}$ in terms of the eigenvalues $\lambda_i$ and eigenvectors of $\underline{\mathbf{A}}$:
\begin{equation}
        f(\underline{\mathbf{A}}) = \sum_{i=1}^n f(\lambda_i) A_i
\label{eq:LSF}
\end{equation}
The solution depends on the dimension. We need to distinguish between 
even and odd dimensions: 
\begin{enumerate}
\item \underline{Even $n$:} $\underline{\mathbf{H}}\in$ Mat$(n\times n$), non-singular, skew-symmetric with 
$n/2$ different eigenvalues.
\item \underline{Odd $n$:} $\underline{\mathbf{H}}$ has one zero-valued eigenvalue and
 the other eigenvalues are
all different. The nonzero eigenvalues appear in $\pm$-pairs and are imaginary.
\end{enumerate}
For the eigenvalue zero, the projector is real and symmetric.
We denote it $\underline{\boldsymbol{\pi}}_0$.
For all other eigenvalues $\mu$, the spectral projectors $\underline{\boldsymbol{\pi}}_{\mu}$ are hermitian matrices. Denote $\sigma$ the set of eigenvalues. 
Now, we apply Lagrange-Sylvester-formula and obtain
\begin{equation}
\exp (t \mathbf{H}) = \sum_{\mu \in \sigma} e^{t \mu} \ \underline{\boldsymbol{\pi}}_{\mu}\, ,
\label{eq:expfun}
\end{equation}
which is the spectral formula for $\exp (t \underline{\mathbf{H}}) $ in terms of the spectral projectors. 
We notice that spectral projectors are orthogonal projections.

\subsubsection{Determining the $(n,n(n+1)/2)$-sub-Riemannian geodesics}

Now, we are prepared to calculate the sub-Riemannian geodesics for the general case of an $n$-dimensional Lie algebra following the work of \citet{Perez2006}. 
The notations $(n, n(n+1)/2)$-sub-Rie\-mannian geometry is composed of a $(n(n+1)/2)$-dimensional Lie algebra with a rank $n$ distribution.
A distribution $\Delta$ of rank $n$ is  a subspace of the tangent space of a manifold, where there exist $n$ linearly independent vector fields that form a basis for the distribution $\Delta$.
Thus, this general case could be adapted to the $n$-dimensional Vortex-Heisenberg group VH(n). 
 The rank of the Lie algebra vh(2) is three (two components of the linear momentum and the circulation) with rank of distribution $n=2$ leading to a (2,3)-sub-Riemannnian structure for two-dimensional incompressible, inviscid flows. 
 The Lie algebra vh(3) of three dimensional vortex flows has six dimensions (6 elements: three components of the linear momentum and three components of the total flux of vorticity) with a rank $n=3$ distribution. 
 Thus, (3,6)-sub-Riemannnian geometry can be applied. For both dimensions, we consider the discretized conservation laws. 
In order to apply their algorithm to two- and three-dimensional atmospheric vortex motions, we will first summarize the algorithm of \citet{Perez2006} to find geodesics for $n$-dimensional algebras. 

In the end of the last section, we suggested to apply the Lagrange-Sylvester  formula \eqref{eq:expfun} to obtain a solution of the following differential equation for geodesics $\mathbf{x}_g$ (see \eqref{eq:normed_curve}):
\begin{equation}
\frac{d\mathbf{x}_g}{dt}  = (\mathbf{\dot{h}}, \dot{ \underline{\mathbf{H}}}).
\end{equation}
To apply the Lagrange-Sylvester formula, we first calculate the eigenspaces of the corresponding systems. 
We will use the following notations for the eigenvectors, eigenvalues and projectors: 
\begin{equation}
\begin{split}
\{ \mathbf{v}_1,\mathbf{v}_{-1},\dots ,  \mathbf{v}_{\lfloor n/2 \rfloor},\mathbf{v}_{-\lfloor n/2 \rfloor} \} \subset \mathbb{C} & \quad \text{(Eigenvectors)}\\
\{i \lambda_1,-i\lambda_1,\dots ,i \lambda_{\lfloor n/2 \rfloor},-i\lambda_{-
\lfloor n/2 \rfloor} \} & \quad  \text{(Eigenvalues)}  \\
 \{ \underline{\boldsymbol{\pi}}_1, \underline{\boldsymbol{\pi}}_{-1},\dots , \underline{\boldsymbol{\pi}}_{\lfloor n/2 \rfloor}, \underline{\boldsymbol{\pi}}_{-\lfloor n/2 \rfloor} \}
& \quad \text{(Projectors)} 
\end{split}
\end{equation}
with $\mathbf{v}_{-k} = \overline{\mathbf{v}}_k$ and $\lambda_i \in \mathbb{R}$. \citet{Perez2006} show that all $\mathbf{v}_i$ are orthogonal, i.e. $\mathbf{v}_i \cdot \mathbf{v}_j = \delta_{ij}$.

We note that for odd \textit{n} we obtain an additional (real) eigenvector $\mathbf{v}_0$, an additional eigenwert $\lambda_0 = 0$ and its projector $\underline{\boldsymbol{\pi}}_0$  $(\mathbf{v}_0 \in {\rm Ker}(H)$).
The orthogonality implies: ${\rm Re}(\mathbf{v}_i)\cdot {\rm Re}(\mathbf{v}_j) = \delta_{ij} {\rm Im}(\mathbf{v}) \cdot {\rm Im}(\mathbf{v}_j)$ and ${\rm Re}(\mathbf{v}_j) \cdot {\rm Im}(\mathbf{v}_j) = 0$. Therefore, 
\begin{equation}
 \{ {\rm Re}(\mathbf{v}_1), {\rm Im}(\mathbf{v}_1), \dots , {\rm Re}(\mathbf{v}_{\lfloor n/2 \rfloor}), {\rm Im}(\mathbf{v}_{\lfloor n/2 \rfloor})\}
\end{equation}
 is an orthogonal basis of $\mathbb{R}^n$ for even $n$. And
\begin{equation}
 \{ {\rm Re}(\mathbf{v}_1), {\rm Im}(\mathbf{v}_1), \dots , {\rm Re}(\mathbf{v}_{\lfloor n/2 \rfloor}), {\rm Im}(\mathbf{v}_{\lfloor n/2 \rfloor}),\mathbf{v}_0 \}
\end{equation}
yields an orthogonal basis of $\mathbb{R}^n$ for odd $n$.
Because $\mathbf{h}_0$ is a real constant vector it is: 
\begin{equation}
  \text{span}({\rm Re}((\mathbf{h}_0 \cdot \mathbf{v}_k) \mathbf{v}_k), {\rm Im}((\mathbf{h}_0 \cdot \mathbf{v}_k) \mathbf{v}_k))
  = \text{span}({\rm Re}(\mathbf{v}_k), {\rm Im}(\mathbf{v}_k) ). 
  \end{equation} 
for $k=1,2, \dots \lfloor n/2 \rfloor$. We obtain a basis
$\{ \boldsymbol{\alpha}_k, \boldsymbol{\beta}_k\}$ with
\begin{equation}
\boxed{
\  \boldsymbol{\alpha}_k = 2\ { \rm Im}((\mathbf{h}_0 \cdot \mathbf{v}_k)\mathbf{v}_k)),\ \boldsymbol{\beta}_k = 2\ {\rm Re}((\mathbf{h}_0 \cdot \mathbf{v}_k)\mathbf{v}_k)), \ \boldsymbol{\gamma}_0 = (\mathbf{h}_0 \cdot \mathbf{v}_0) \mathbf{v}_0 .\ }
\label{eq:basis}
\end{equation}
for $k=1,2, \dots \lfloor n/2 \rfloor$.

We search for the solution for the pair $ (\mathbf{x}_g, \underline{\mathbf{z}}_g)\in \mathbb{R}^n \times \mathfrak{so}(n)$, i.e. the geodesic arc. Let this geodesic arc be defined in a certain interval with the following initial condition 
\begin{equation}
(\mathbf{x}_g(0), \underline{\mathbf{z}}_g(0))= (\mathbf{0},\underline{\mathbf{0}}).
\label{eq:incond}
\end{equation}
Assuming that $(\mathbf{x}_g,  \underline{ \mathbf{z}}_g)$ is a projection of a normal extremal and all eigenvalues of $\underline{\mathbf{H}}$ are non-zero, we can apply (\ref{eq:expfun}) leading to the following geodesic equations:
\begin{equation}
\mathbf{x}_g = 
\begin{cases} 
\sum_{\mu \in (\sigma)} \frac{1}{\mu} (e^{\mu t}-1) \ \underline{\boldsymbol{\pi}}_{\mu} \mathbf{v}_0 & \text{for n even} \\
\sum_{\mu \in (\sigma-\{0\})} \frac{1}{\mu} (e^{\mu t}-1) \ \underline{\boldsymbol{\pi}}_{\mu} \mathbf{v}_0+ t \underline{\boldsymbol{\pi}}_0 \mathbf{v}_0 & \text{for n odd} 
\end{cases}
\end{equation}
Now, we can formulate the geodesic arc $ (\mathbf{x}_g,\mathbf{z}_g)$ with respect to the above basis \eqref{eq:basis} given by $\{\boldsymbol{\alpha}_k,\boldsymbol{\beta}_k  \}$ for $n$ even, and by $\{\boldsymbol{\alpha}_k,\boldsymbol{\beta}_k, \boldsymbol{\gamma}_0  \}$ for $n$ odd. We obtain the trajectories:
\begin{equation}
 \boxed{
\begin{split}
\mathbf{x}_g & = \sum_{i=1}^{\lfloor n/2 \rfloor} \frac{1}{\lambda_i} (\cos(\lambda_i t)-1)  \boldsymbol{\alpha}_i
+ \frac{1}{\lambda_i} \sin(\lambda_i t ) \boldsymbol{\beta}_i \hspace{1.3cm} \text{ (for $n$ even)}\\
\underline{\mathbf{z}}_g & = \sum_{i=1}^{\lfloor n/2 \rfloor} \ A_{ij} \boldsymbol{\alpha} \wedge 
\boldsymbol{\alpha}_j
+ B_{ij} \boldsymbol{\alpha}\wedge \boldsymbol{\beta}_j
+ C_{ij} \boldsymbol{\beta} \wedge \boldsymbol{\beta}_j\qquad \text{ (for $n$ even)}
\end{split}
}
\label{eq:TrajecSubrgen}
\end{equation}
with
\begin{equation}
\begin{split}
A_{ij} & = \frac{1-\cos(\lambda_i- \lambda_j) t}{2 \lambda_i (\lambda_i - \lambda_j)}  
+ \frac{\cos(\lambda_i + \lambda_j ) t-1)}{2 \lambda_i (\lambda_i - \lambda_j)} 
- \frac{\cos(\lambda_j t) - 1}{\lambda_i \lambda_j} \\
B_{ij} & = \frac{(\lambda_j - \lambda_i )\sin(\lambda_i + \lambda_j)t }{2 \lambda_i \lambda_j  (\lambda_i+ \lambda_j)}
+ \frac{(\lambda_j +\lambda_i )\sin(\lambda_i- \lambda_j)t }{2 \lambda_i \lambda_j  (\lambda_i- \lambda_j)}
- \frac{\sin(\lambda_j t)}{\lambda_i \lambda_j},\  i \neq i, 
\end{split}
\end{equation} 
\begin{equation}
\begin{split}
B_{ii} & = \frac{t}{\lambda_i} - \frac{\sin(\lambda t)}{\lambda_i^2} \\
C_{ij} & = \frac{1-\cos(\lambda_i- \lambda_j ) t }{2 \lambda_i (\lambda_i - \lambda_j)}
- \frac{\cos(\lambda_i+\lambda_j)t -1}{2 \lambda (\lambda_i +\lambda_j)}.
\end{split}
\end{equation} 
For $n$ odd the same equations \eqref{eq:TrajecSubrgen} hold but with the additional
$t\boldsymbol{\gamma}_0$ and with the additional two terms:
\begin{equation}
\begin{split}
\sum_{i=1}^{\lfloor n/2 \rfloor} & \left( 
\frac{-t}{\lambda_i} (\cos(\lambda_it) + 1) + \frac{1}{\lambda_i^2} 2 \sin (\lambda_i t) \boldsymbol{\alpha_i} \wedge \boldsymbol{\gamma}_0
 \right) \\
 & + \sum_{i=1}^{\lfloor n/2 \rfloor}   
 \left( \frac{1}{\lambda_i^2} (2(1 - \cos(\lambda_i t))- \lambda_i t \sin(\lambda_i t)) 
 \right)
 \boldsymbol{\beta_i} \wedge \boldsymbol{\gamma}_0
 \end{split}
\end{equation}
We cite \citet{Perez2006} and summarize that in even dimensions the projections of the components of $\mathbf{x}_g$ to the planes $\{{\rm Re}(\mathbf{v}_k), {\rm Im}(\mathbf{v}_k) \}$, for $k=1,2, \dots \lfloor n/2 \rfloor$,  are circles passing through the origin with radii $1/ \lambda_k$, and centered at span$( \boldsymbol{\alpha}_k)$. 
Furthermore, in odd dimensions, since $\mathbf{x}_g$ varies linearly in the direction 
 of the vector $\boldsymbol{\gamma}_0$, the projections of the component $\mathbf{x}_g$ to the three-dimensional
 subspaces span$( \boldsymbol{\alpha}_k, \boldsymbol{\beta}_k,\boldsymbol{\gamma}_0)$ (identical to span$({\rm Re}(\mathbf{v}_k), {\rm Im}(\mathbf{v}_k))$) are helices. In this case we can write explicitly the parameter $t = (\mathbf{h}_0 \cdot (\underline{\boldsymbol{\pi}}_0 \mathbf{x}_g))/\Vert \underline{\boldsymbol{\pi}}_0 \mathbf{h}_0 \Vert^2$
 %\item Perez: According to Theorem 4.2, a geodesic arc which is a projection of an abnormal extremal (not strictly abnormal) corresponds to a straight line. 
 %Since in the case $\gamma_0 \in Ker(H)$, we can conclude that for odd n, the helix idescribed in the previous remark, which is a projection of a normal extremal, has a directrix the eigenspace corresponding to the eigenvalue zero, which coincides with the line given by the abnormal extremal. 

\section[(2,3)-sub-Riemannian geometry]{(2,3)-sub-Riemannian geometry}
\label{sec:2.3allg} 

In the last section we have summarized the algorithm the general $(n,n(n+1)/2))$-sub-\-Riemannian geodesics for a $n(n+1)/2$-dimensional Lie group with a rank $n$ distribution. This algorithm could be applied to the $n$-dimen\-sional Vortex-Heisenberg group, which might be explored in future studies. Here, we regard the  special case of 
%The nilpotent  holds a natural (2,3)-sub-Riemannian structure. 
(2,3)-sub-Riemannian geometry, which was introduced by \citet{Brockett1982}, 
to find geodesics for the two-dimensional vortex flows. 
A classical example is given by charged mass points, \citep[see, e.g.,][]{Monroy1999}. Its algebraic structure is isomorphic to the two-dimensional Vortex-Heisenberg algebra.

Here, we will first outline the derivation of the general (2,3)-\-sub-\-Rie\-mannian geodesic after \cite{Perez2006} in order to apply this algorithm to find geodesics for the idealized point vortex model.
Physically, the point vortex motion is restricted by the conservation laws, and all of them can be expressed with respect to the circulation. Mathematically, the conservation laws imply a nilpotent Lie algebra such that sub-Riemannian geometry is a suitable choice to find vortex geodesics. 

Consider the following nonzero bracket of the three-dimensional, se\-cond-step nilpotent Lie Algebra:
\begin{equation}
[\mathbf{X}_1,\mathbf{X}_2] = \mathbf{X}_{12}.
\end{equation} 
We use the same notations as in the last section for arbitrary $n \in \mathbb{N}$ and denote with $H_i$ the Hamiltonian functions corresponding to the vector field $\mathbf{X}_i$, $i=1,2$, respectively $H_{ij}$ to $\mathbf{X}_{ij}$, $i,j=1,2$. 
Then, we obtain the  pair $(\mathbf{h}, \underline{\mathbf{H}})\in \mathbb{R}^2 \times \mathfrak{so}(2)$ (see \eqref{eq:pair}).
Now, we consider the following initial values 
\begin{equation}
\mathbf{h}_0 = (H_1,H_2)^T .
\end{equation}
and solve the equation
 \begin{equation}
\frac{d\mathbf{x}_g}{dt}  = (\mathbf{\dot{h}}, \dot{ \underline{\mathbf{H}}}) 
\label{eq:geod2D}
\end{equation}
to derive the vortex geodesics $\mathbf{x}_g$.

In order to solve \eqref{eq:geod2D}, we apply Lagrange-Sylvester formula (\ref{eq:LSF}) leading to two eigenvalues 
\begin{equation}
\{- i \lambda_1, i \lambda_1 \}
\end{equation} 
with $\underline{\mathbf{H}}^2+\lambda_1^2 \mathbf{E} = 0$, where $\underline{\mathbf{E}}$ denotes the $2\times 2$ identity matrix.
The spectral projector is given by 
\begin{equation} 
\underline{\boldsymbol{\pi}}_1 = \frac{1}{2i \lambda_1} (\underline{\mathbf{H}}+i \lambda_1 \underline{\mathbf{E}}).
\end{equation}
And, using (\ref{eq:basis}), the basis of the projection plane reads as
\begin{equation}
\boldsymbol{\beta}_1 = (H_1,H_2)^T,\  \boldsymbol{\alpha}_1 = (-H_2,H_1)^T.
\end{equation}
Finally, the geodesic equation is given by: 
\begin{equation}
\boxed{
\mathbf{x}_g  = 
\frac{1}{\lambda_1} (\cos(\lambda_1 t)-1) \boldsymbol{\alpha}_1 + \frac{1}{\lambda_1} \sin(\lambda_1 t) \boldsymbol{\beta}_1 
}
\end{equation}
and
\begin{equation}
\underline{\mathbf{z}}_g  = 
\left( \frac{t}{\lambda_1} - \frac{\sin(\lambda_1 t)}{\lambda_1^2}\  \right)\boldsymbol{\alpha}_1 \wedge \boldsymbol{\beta}_1,
\end{equation}
where $\wedge$ denotes the wedge-product.

\section{Deriving geodesics for point vortex systems}
\label{subsec:2-3subr}

We consider the discretized, idealized point vortex model, which hold the analogous bracket relations as vh(2). 
The first formulations of the equations of point vortices in the plane  can be ascribed to \citet{Helmholtz1858}.
Twenty years later, \citet{Kirchhoff1876} introduced the general Hamiltonian structure of $N$ point vortices, shortly followed by
\citet{Groebli1877}, who analysed in detail the motion of three point vortices, see also e.g. \citet{Synge1949}, \citet{Novikov1975}, \citet{Aref1979}, \citet{Obukhov1984},\citet{Makhaldiani1998}, \citet{Newton2001}, \citet{Aref2007}, \citet{Blackmore2007}, or \citep{Chapman1978} who derived the point vortex equations via a variational principle.

Denote $\mathbf{x}_i= (x_i \ y_i)^T , \ i=1,\dots N$ the local coordinates of the $i$-th 
point vortex of a $N$-point vortex system in the plane. Each vortex is characterized by its circulation $\Gamma_i$, $i=1,2,\dots ,N$, which characterizes the strength and the direction of rotation. 
Further, denote $r_{ij}= ( (x_i-x_j)^2 + (y_i -y_j ) ^2)^{1/2}$ the relative distance of 
the $i$-th and $j$-th point vortex ($i,j=1,\dots ,N$).
Then, the equations of motion derived by \citet{Helmholtz1858} are given by:
\begin{equation}
\begin{split}
\frac{d x_j }{d t}  = - \frac{1}{2 \pi} \sum_{i\neq j \atop {i,j=1}}^N\frac{\Gamma_i (y_j-y_i ) }{r_{ij}^2}, \quad
\frac{d y_j }{d t} = + \frac{1}{2 \pi} \sum_{i\neq j \atop {i,j=1}}^N \frac{\Gamma_i (x_j-x_i ) }{r_{ij}^2} .
\label{eq:bew}
\end{split}
\end{equation}
We remark that \citet{Kirchhoff1876} established the Hamiltonian representation of these equations of motion as non-linear coupled system of $2N$ ordinary differential equations.
The Nambu representation of point vortex dynamics is discussed in \citet{Mueller2014}. 
The zonal and meridional momenta and the total circulation for a $N$-point vortex system are conserved quantities given by: 
\begin{equation}
P_x = \sum_{i=1}^{N} \Gamma_i y_i,\quad
P_y = - \sum_{i=1}^{N} \Gamma_i x_i 
\quad \text{and} \quad 
\Gamma =  \sum_{i=1}^{N} \Gamma_i \, .
\end{equation}

The composition of $P_x,P_y$ and
$\Gamma$ leads to a further important conserved quantity called the \textit{center of circulation}
$\mathbf{C} $: 
\begin{eqnarray}
\mathbf{C} = \frac{ \sum_i^N \Gamma_i \boldsymbol{x}_i }{ \sum_i^N \Gamma_i}\, .
\label{eq:centre}
\end{eqnarray}
Regarding a three point vortex system there are three kinds of motions: 
(i) periodic motion around the center of circlulation, (ii) collapse/expanding of the three vortices, still rotating around the center of circulation, and (iii) equilibrium, where the vortices form an equilateral triangle.

%%%%%%
%%%%%%% einfuegen ende
\citet{Nevir1998} shows that the bracket relations of discrete, two-dimensional point vortex systems are equivalent to the Nambu bracket relations of continuous, two dimensional, incompressible, inviscid flows with respect to the enstrophy: 
\begin{equation}
[P_x,P_y] = \Gamma , \quad \text{and} \quad 
[P_x,\Gamma] = [P_y,\Gamma] = 0.
\end{equation}

We consider the single trajectories of the $i$-th point vortex of an $N$-point vortex system and show examples of two and three point vortex systems. 
First, we scale the momentum of the $i$-th point vortex $\mathbf{P}^{(i)}= (P_x^{(i)}, P_y^{(i)})$  by the area $F$:
\begin{equation}
P_x^{(i)} = \frac{\Gamma}{F} y, \qquad P_y^{(i)} = -\ \frac{\Gamma}{F} x
\end{equation}
such that the unit of $P_x$ and $P_y$ is $m/s$, a velocity.  
We recall that $\Gamma$ denotes the circulation. 
W.o.l.g. we assume that $\left(\frac{\Gamma}{F}\right)^2 = 1$.
The $(2 \times 2)$-matrix $\underline{\mathbf{H}}$ and the vector $\mathbf{h} \in \mathbb{R}^2$ form the pair $(\mathbf{h},\underline{\mathbf{H}})$ which is given by:
\begin{equation}
\underline{\mathbf{H}} =  \begin{pmatrix} \{P_x^{(i)},P_x^{(i)} \} & \{P_x^{(i)},P_y^{(i)}\} \\ \{P_y^{(i)},P_x^{(i)}\} & \{P_y^{(i)},P_y^{(i)}\}  \end{pmatrix} = 
 \begin{pmatrix} 0 & \frac{\Gamma}{F}\\ -\frac{\Gamma}{F} & 0 \end{pmatrix},\quad 
\mathbf{h} = \begin{pmatrix} H_1 \\ H_2 \end{pmatrix}
 = \begin{pmatrix} P_x^{(i)} \\ P_y^{(i)} \end{pmatrix}
 \label{eq:HandhPW}
\end{equation}
with respect to the Nambu-bracket $\{\cdot,\cdot \}$.

We recall that the geodesics are the solution of the following differential equation that was derived in \eqref{eq:normed_curve}
\begin{equation}
 \frac{d\mathbf{x}_g}{dt}  = (\mathbf{\dot{h}}, \dot{ \underline{\mathbf{H}}}) = ( \underline{\mathbf{H}}\mathbf{h},\mathbf{0}).
 \label{eq:geodH}
\end{equation}
Here, the components of $\mathbf{h}$ are the linear momenta of the single vortices that form together a $N$-point vortex system.
In contrast to classical mechanical systems, where the momentum is given by the velocity, here, the linear momenta depend on the local coordinates  -- one hierarchical level lower than the mass point momenta. 
The components of $\mathbf{h}$ are given by the linear momenta, and thus, $\mathbf{h}$ depends on the local coordinates $\mathbf{h} = \mathbf{h}(x,y)$ .
Therefore, for the general case, where we do not consider the origin, we can assume $\mathbf{\dot{h}} \neq 0$. On the other hand, $\underline{\mathbf{H}}$ is formed by the circulations that are constant, therefore, it is $\dot{ \underline{\mathbf{H}}}=0$.

In order to solve \eqref{eq:geodH}, we apply Lagrange-Sylvester-formula, i.e. the spectral formula in terms of spectral projectors. To derive the eigenvalues, eigenvectors and projectors, we first calculate the characteristic polynomial of the rotational matrix $\underline{\mathbf{H}}$:
\begin{equation}
\begin{split}
 \det(\underline{\mathbf{H}}-\kappa \underline{\mathbf{E}}) 
 & = \det\left( \begin{pmatrix} -\kappa & \frac{\Gamma}{F} \\ -\frac{\Gamma}{F} & -\kappa \end{pmatrix} \right) 
  = \kappa^2+ \left( \frac{\Gamma}{F} \right) ^2 \\
 &  =  \left( \kappa + i \frac{\Gamma}{F} \right)\ \left( \kappa- i \frac{\Gamma}{F} \right) 
  = 0 .
  \end{split}
\end{equation}
Thus, the solutions $\kappa_1$ and $\kappa_2$ of last equation are the eigenvalues:
\begin{equation}
\kappa_1 =  i \frac{\Gamma}{F}, \quad \kappa_2 =  -i \frac{\Gamma}{F}\quad \Rightarrow \quad \lambda = \frac{\Gamma}{F}
\end{equation}
We will use the following notation:
\begin{equation}
\{\kappa_1, \kappa_2 \} 
= \{i \lambda, - i \lambda\} 
= \left\{ i \frac{\Gamma}{F}, - i \frac{\Gamma}{F}\right\}
\end{equation}
The algebraic multiplicity  $m_k$ of both eigenvalues is one. 
Further, consider the hauptspace 
\begin{equation}
{\rm Haupt}(\kappa_k,\underline{\mathbf{H}}) = {\rm Kern}(\underline{\mathbf{H}}-\kappa_k \underline{\mathbf{E}})^{m_k}
\end{equation}
of the eigenvalue $\kappa_k$. Denote $\underline{\mathbf{B}}$ the block matrix of eigenvectors.
Thus, $\underline{\mathbf{B}}$ is invertible and given by
\begin{equation}
\underbrace{\underline{\mathbf{B}}}_{n \times n} := \left( \underbrace{\mathbf{v}_1}_{n\times  m_1} \vert \dots \vert \underbrace{\mathbf{v}_k}_{n \times  m_n} \right).
\end{equation}
In our example it is $n=2$ and the algebraic multiplicity of both eigenvalues is one, i.e., $m_1 = m_2 = 1$, we summarize:
\begin{equation}
 \kappa_1 =+i\ \frac{\Gamma}{F},  \kappa_2 = -i\ \frac{\Gamma}{F},\ \Rightarrow \lambda = \frac{\Gamma}{F}, \
\mathbf{v}_1 = \begin{pmatrix} i \\  1 \end{pmatrix}, \ 
\mathbf{v}_2 =\begin{pmatrix} -i \\ 1 \end{pmatrix},
\end{equation}
where $\mathbf{v}_1$ and $\mathbf{v}_2$ are the eigenvectors. 
The geometric mulplicity here must be equal to the algebraic multiplicity, because the algebraic multiplicity is one and the geometric mulplicity is smaller than (or equal to) the algebraic mulplicity and larger than (or equal to) one. Therefore, the hauptspace is equals to the eigenspace and thus the basis of the hauptspace can be represented by the above introduced matrix $\underline{\mathbf{B}}$ given by
\begin{equation}
\underline{\mathbf{B}} = (\mathbf{v}_1 \vert \mathbf{v}_2) = \begin{pmatrix} i  &  -i \\ 1  &  1 \end{pmatrix} 
\quad \text{and} \quad
\underline{\mathbf{B}}^{-1} = \frac{1}{2i}  \begin{pmatrix} 1 & i \\  -1 & i \end{pmatrix} =: \begin{pmatrix} \mathbf{C}_1 \\ \mathbf{C}_2\end{pmatrix}.
\end{equation}
Now, we can determine the spectral projectors:
\begin{equation}\begin{split}
\underline{\boldsymbol{\pi}}_1 &= \mathbf{v}_1\ \mathbf{C}_1 
= \frac{1}{2i} 
\begin{pmatrix} i \\  1 \end{pmatrix}  
\begin{pmatrix} 1 & i \end{pmatrix}
= \frac{1}{2i} 
\begin{pmatrix} i & -1 \\ 1 & i \end{pmatrix} \\ 
\underline{\boldsymbol{\pi}}_2 &= \mathbf{v}_2\ \mathbf{C}_2 
= \frac{1}{2i} 
\begin{pmatrix} -i \\  1 \end{pmatrix}  
\begin{pmatrix} -1 &  i \end{pmatrix}
= \frac{1}{2i} 
\begin{pmatrix} i & 1 \\  -1 & i \end{pmatrix}
\end{split}
\end{equation}
and show that:
\begin{equation}
\underline{\mathbf{H}}^2 +\kappa^2 \underline{\mathbf{E} }
= \underline{\mathbf{H}}^2 + \left( \frac{\Gamma}{F}\right)^2 \underline{\mathbf{E}}
= \begin{pmatrix} 0 &  \frac{\Gamma}{F} \\ -  \frac{\Gamma}{F} & 0 \end{pmatrix}^2
+\begin{pmatrix}  \frac{\Gamma}{F} & 0 \\ 0&  \frac{\Gamma}{F} \end{pmatrix} 
= \begin{pmatrix} 0 & 0 \\ 0& 0\end{pmatrix} .
\end{equation}
%We obtain:
%\begin{equation}
%\frac{1}{2i \kappa} (\mathbf{H}+i\lambda \mathbf{E}) 
%= \frac{1}{2i \Gamma / F} \left(\mathbf{H}+i \frac{\Gamma}{F} \ \mathbf{E} \right) 
%= \frac{F}{2i \Gamma } \begin{pmatrix} i  \frac{\Gamma}{F} &  \frac{\Gamma}{F} \\ - \frac{\Gamma}{F} & i \frac{\Gamma}{F} \end{pmatrix} .
%\end{equation}
Then, with (\ref{eq:basis}) the basis vectors are given by 
\begin{equation}
\boldsymbol{\alpha}^{(i)} = \begin{pmatrix} -H_2\\ H_1\end{pmatrix} = \begin{pmatrix} -P_y^{(i)}\\ P_x^{(i)}\end{pmatrix},\ 
\boldsymbol{\beta}^{(i)} =\begin{pmatrix} H_1\\ H_2\end{pmatrix} =  \begin{pmatrix} P_x^{(i)}\\ P_y^{(i)}\end{pmatrix}.
\end{equation}
Inserting $\boldsymbol{\alpha}, \boldsymbol{\beta}$ and $\lambda$ into the following geodesic equation (see \eqref{eq:TrajecSubrgen}):
\begin{equation}
\mathbf{x}_g^{(i)}  = \frac{1}{\lambda} \left( \cos(\lambda t)-1 \right) \boldsymbol{\alpha}^{(i)}
+ \frac{1}{\lambda} \sin(\lambda t) \boldsymbol{\beta} ^{(i)}
\end{equation}
leads to the sub-Riemannian geodesics for the $i$-th point vortex of an $N$-point vortex system: %in an two-dimensional incompressible, inviscid vortex flow:
\begin{equation}
\boxed{
 \mathbf{x}_g^{(i)} = \frac{F}{\Gamma}  \left( \cos \left(\frac{\Gamma}{F} t \right)-1\right) \begin{pmatrix} -P_y^{(i)}  \\ P_x^{(i)}\\ \end{pmatrix}
+ \frac{F}{\Gamma } \sin \left(\frac{\Gamma}{F}  t \right) \begin{pmatrix} P_x^{(i)}  \\ P_y^{(i)}\\ \end{pmatrix}
\label{geod}
}
\end{equation}
with respect to the phase space with total circulation $\Gamma$, linear momenta $P_x^{(i)}$ and $P_y^{(i)}$ and a vortex-surface-parameter $F$. 
Here we considered $N$-point vortex systems. 
We hypothesize that sub-Riemannian geodesics in $N$-point vortex systems can be regarded as relative equilibria solutions of the point vortex equations of motions. 
We focus on $N=2$- and $N=3$-point vortex systems, because point vortex systems formed by one, two or three point vortices are integrable. 
Vortex constellations of more than three point vortices will be investigated in future studies.

%--------------------------------------------------------

\subsection[]{Comparing the sub-Riemannian-geodesics to point vortex trajectories for $N=2$ and $N=3$} 
 
\begin{figure}[t]
\centering{
\includegraphics[scale=.7]{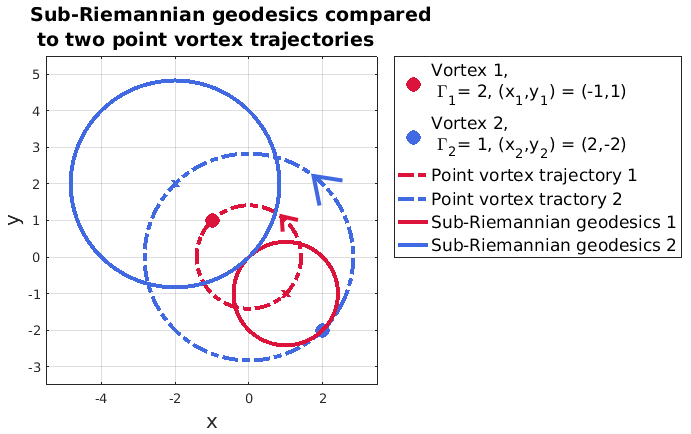} %\includegraphics[scale=.8]{2Wirbel_2D.eps}
\caption{\small{The solid line are the trajectories of the sub-Riemannian geodesics of two vortices, whereas the dashed line are the point vortex trajectories. 
The circulations of the systems where chosen to be equal but the trajectories are calculated with different models. Their trajectories are congruent.}}
\label{fig:2DSubrie2V}
}
\end{figure}

In order to compare the (2,3)-sub-Riemannian geodesics \eqref{geod} to point vortex dynamics we reformulate last expression \eqref{geod} for the geodesic $\mathbf{x}_g^{(i)}$:
\begin{equation}
\begin{split}
\mathbf{x}_g^{(i)} & = \frac{F}{\Gamma} \left( \cos \left(\frac{\Gamma}{F} t \right)-1\right) 
\begin{pmatrix} -P_{y}^{(i)}  \\ P_{x}^{(i)}\\ \end{pmatrix}
+ \frac{F}{\Gamma} \sin \left( \frac{\Gamma}{F} t \right) \begin{pmatrix} P_{x}^{(i)} \\ P_{y}^{(i)}\\ \end{pmatrix} \\
 & = \frac{F}{\Gamma} \left( \cos \left(\frac{\Gamma}{F} t \right) -1 \right) 
 \begin{pmatrix} \frac{\Gamma}{F} x  \\  \frac{\Gamma}{F} y \\ \end{pmatrix}
+ \frac{F}{\Gamma} \sin \left(\frac{\Gamma}{F} t \right) 
\begin{pmatrix} \frac{\Gamma}{F}  y  \\  -\frac{\Gamma}{F} i x  \\ \end{pmatrix} \\
 & = \frac{F}{\Gamma} 
 \begin{pmatrix} \cos \left( \frac{\Gamma}{F} t \right) \frac{\Gamma}{F} x - \frac{\Gamma}{F} x
 + \sin \left( \frac{\Gamma}{F} t \right) \frac{\Gamma}{F} y \\ & \\
  \cos \left(\frac{\Gamma}{F} t \right) \frac{\Gamma}{F} y - \frac{\Gamma}{F} y
  - \sin \left(\frac{\Gamma}{F} t \right) \frac{\Gamma}{F} x
 \end{pmatrix} \\
 & = \frac{\Gamma}{F}\frac{F}{\Gamma}
  \begin{pmatrix}  
     \cos \left(\frac{\Gamma}{F} t \right) x + \sin \left( \frac{\Gamma}{F} t \right)  y \\ & \\
  - \sin \left( \frac{\Gamma}{F} t \right)  x + \cos \left( \frac{\Gamma}{F} t \right)  y 
 \end{pmatrix} 
 - \frac{\Gamma}{F}\frac{F}{\Gamma}
 \begin{pmatrix}  x \\  y \end{pmatrix} \\
 & =\begin{pmatrix}  
     \cos \left( \frac{\Gamma}{F} t \right)  &  \sin \left(\frac{\Gamma}{F} t \right)   \\ & \\
  - \sin \left( \frac{\Gamma}{F} t \right) &  \cos \left( \frac{\Gamma}{F} t \right)   
 \end{pmatrix} 
\begin{pmatrix}  x \\  y
 \end{pmatrix} 
 - \begin{pmatrix}  x \\  y
 \end{pmatrix}.
\end{split}
\end{equation}
Now, we set
\begin{equation}
 \underline{\mathbf{R}} \left(\frac{\Gamma}{F} t \right) :=
\begin{pmatrix}  
     \cos(\frac{\Gamma}{F} t)  &  \sin(\frac{\Gamma}{F} t)   \\
    & \\
  - \sin(\frac{\Gamma}{F} t) &  \cos(\frac{\Gamma}{F} t)   
 \end{pmatrix} .
\label{eq:subriem2drot1}
\end{equation}
Thus, we can express the geodesic equation:
\begin{equation}
\boxed{\mathbf{x}_g^{(i)} = \underline{\mathbf{R}} \left( \frac{\Gamma}{F} t \right)\ \mathbf{x} - \mathbf{x} }
\label{eq:subriem2drot2}
\end{equation}
Therefore, vortex geodesics in the phase space spanned by $\boldsymbol{\alpha}$ and $\boldsymbol{\beta}$ are rotations around the  initial coordinates that are reflected. The geodesics pass the origin as shown by the solid lines in fig. \ref{fig:2DSubrie2V}.

\begin{figure}[t]
\includegraphics[scale=.7]{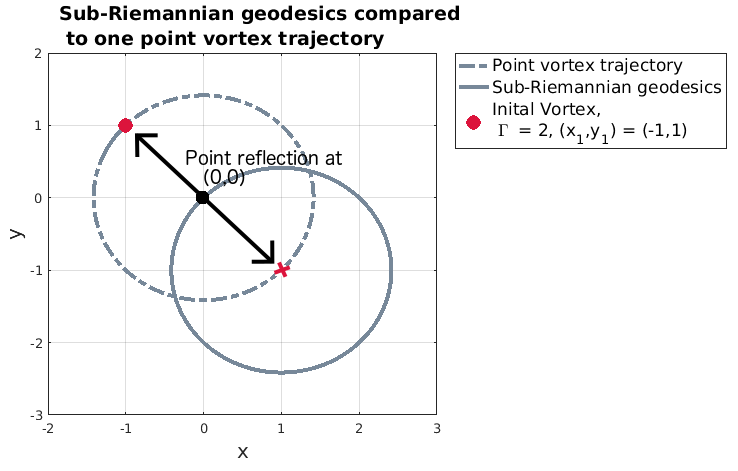} %\includegraphics[scale=.8]{3Wirbel_eqilibrium.eps}
\caption{\small{The colored trajectories show the sub-Riemannian geodesic of three vortices based on the Vortex-Heisenberg Lie algebra. The black line illustrates the point vortex motion. Even though both trajectories are determined with different models, we see that both trajectories are congruent.}}
\label{fig:2DSubrie}
\end{figure}

Let us now compare this result with a point vortex system.
One point vortex systems always remains in calm, because they are located in its center of circulation. 
Thus, we first consider the simplest non-trivial dynamical system of two point vortices. 
Unless their total circulation is equal to zero, two vortices rotate uniformly around the center of circulation \eqref{eq:centre}. 
See e.g. fig. \ref{fig:2DSubrie2V}, where the center of circulation is in the origin and the red and blue point vortex trajectories are shown by the dashed lines. The dots show the initial coordinates of the two point vortices and the arrows indicate the direction of the rotation. 
Here, the total circulation (the sum of the circulations of the single vortices) is three. For systems with total circulation equal zero the center of circulation would approach infinity and the whole point vortex system would translate. 
%In fig. \ref{fig:2DSubrie} the center of circulation lies at the origin and the point vortex motion in the $x$-$y$-plane is indicated by the gray dashed circle. 
To illustrate the point vortex trajectories the point vortex equations \eqref{eq:bew} were calculated using MATLAB's ode45 solver that is based on an explicit Runge-Kutta formula.
% which is called relative equilibrium if the three vorticies forms an equilateral triangle. 
%See e.g. the three point vortices marked by the colored dots in fig. \ref{fig:2DSubrie}. 
%In this example, the total circulation is unequal to zero, 
%$\Gamma_1 +  \Gamma_2 + \Gamma_1= 3 \neq 0$. 
%Three vortex systems that form an equilateral triangle and have non-vanishing total circulations are called equilibrium, as we have discussed in chapters \ref{Nambu:PW} and \ref{chap:Blockings}. 
%The colored paths indicate the sub-Riemannian geodesics \eqref{geod} with respect to the basis $\boldsymbol{\alpha}$ and $\boldsymbol{\beta}$, where each sub-Riemannian vortex trajectory is regarded separately.
%In fig. \ref{fig:2DSubrie}, all three sub-Riemannian trajectories pass the origin.
%This is due to the initial conditions (see \eqref{eq:incond}), where the vortices are shifted to the origin. Therefore, each sub-Riemannian trajectory is shifted. %Furthermore, the sub-Riemannian geodesics are rotations, where the center is the at the origin reflected local coordinate of the initial point vortices, as illustrated in fig. \ref{fig:2DSubrie}. 
The colored solid lines in fig. \ref{fig:2DSubrie} indicate the corresponding sub-Riemannian geodesics with respect to the basis $\boldsymbol{\alpha}$ and $\boldsymbol{\beta}$. 
Their projection would lead to a representation in the space of local coordinates. 
%they are only illustrated in the same figure as the point vortex equilibrium to indicate the differences and congruence of both systems, they 'live' on a different plane. 
Regarding \eqref{eq:subriem2drot1} and \eqref{eq:subriem2drot2}, the center of the sub-Riemannian geodesics are always given by the coordinate that results from the point reflection of the initial coordinate of the point vortices. 
This is due to the initial conditions given by the starting point $(0,0)$, because the Vortex-Heisenberg group identity is in $(0,0)$ and the Vortex-Heisenberg Lie algebra is the tangent space at the group identity. The centers of the sub-Riemannian geodesics are indicated by the crosses in fig. \ref{fig:2DSubrie2V} and fig. \ref{fig:2DSubrie}. 
The rotational centers of the sub-Riemannian geodesics lie on the point vortex trajectory.
Thus, the point vortex trajectory and the sub-Riemannian geodesics have the same radius! 

%%% SKIZZE

\begin{figure}[t]
\includegraphics[scale=.65]{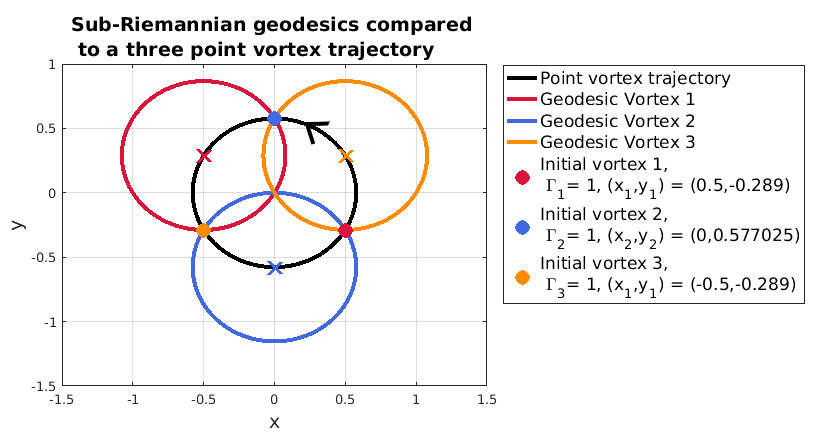} %\includegraphics[scale=.8]{3Wirbel_eqilibrium.eps}
\caption{\small{The colored trajectories show the sub-Riemannian geodesics of three vortices based on the Vortex-Heisenberg Lie algebra. The black line illustrates the  motion of the three-point vortex system forming an relative equilibrium. 
The black point vortex trajectory was determined by solving the the point vortex equations directly. Even though the trajectories are determined with different approaches, we see that the sub-Riemann geodesics and the point vortex trajectories are congruent.}}
\label{fig:Subr3W}
\end{figure}

As second example we consider a three point vortex system. 
It is called relative equilibrium if the three vortices forms an equilateral triangle. One example of a three point vortex equilibrium is shown in fig. \ref{fig:Subr3W}, where the initial locations of the vortices are marked by the colored dots.
In this example, the total circulation is unequal to zero, 
$\Gamma_1 +  \Gamma_2 + \Gamma_1= 3 \neq 0$. 
Three point vortex equilibria rotate uniformly around the center of circulation \eqref{eq:centre}. If the total circulation is zero, the center of circulation would approach infinity and the whole point vortex system would translate. 
In fig. \ref{fig:Subr3W} the center of circulation lies at the origin and the point vortex motion in the $x$-$y$-plane is indicated by the black circle. 
Here, the equations of motion \eqref{eq:bew} were solved to illustrate the point vortex equilibrium motion using MATLAB's ode45 solver that is based on an explicit Runge-Kutta (4,5) formula.
The colored solid lines in fig. \ref{fig:Subr3W} show the sub-Riemannian geodesics \eqref{geod} with respect to the basis $\boldsymbol{\alpha}$ and $\boldsymbol{\beta}$, where each sub-Riemannian vortex trajectory is regarded separately with the common total circulation $\Gamma$.  
All trajectories pass the Vortex-Heisenberg group identity element $(0,0)$ which were chosen as initial condition for the general sub-Rieman\-nian geo\-desics (see \eqref{eq:incond}), because the Lie algebra is the tangent space at the identity element. 
Therefore, compared to the point vortex trajectory, each sub-Riemannian trajectory is shifted to pass the origin.
The black line shows the point vortex motion of the equilibrium. 
As we found for the two-vortex systems, the sub-Riemannian geodesics are rotations, where the center is given by the point reflection of the initial coordinates, see fig. \ref{fig:Subr3W}, where the centers are marked by the colored crosses.
They lie on the point vortex trajectory. 
We can conclude that also for the three-vortex equilibrium the single point vortex trajectories and the sub-Riemannian geodesics have the same radius! 

We have shown examples of two-and three-point vortex systems, where the sub-Riemannian geodesics and the point vortex trajectories are congruent. 
In both examples, the center of circulation of the point vortex systems lies at the origin. If one choses different center of circulation, the trajectories become scaled, and a scaling factor need to be respected. 
If the total sum of circulation is equal to zero, the point vortex system equilibrium would translate. This also coincides with sub-Riemannian geodesics, because in this case, we obtain an abnormal extremal which, independent of the applied physical realization, leads to a straight line as geodesics \citep[see, e.g.][]{Perez2006}. 
A vortex is characterized by its vortical rotation. All physical conserved quantities contain the circulation. 
Therefore, it is natural that -- in contrast to mass point dynamics -- point vortices usually do not move on straight lines.

Why are point vortex geodesics congruent with point vortex equilibria trajectories?
Let us regard one point vortex of a two-or three point vortex system, assuming that the vortex system does not form an equilibrium. We also assume that the vortex does not expand or collapse.
In this case, the vortex interacts with other point vortices and rotates several times around the center of circulation until it arrives its starting point again. 
Sometimes, a strong interaction between the vortices leads to additional rotations around another point vortex. 
Now, we measure and compare the lengths of the paths of the vortex that rotates several times around the center of circulation and interacts with other vortices of the point vortex system. We find that a point vortex that is part of an equilibrium takes the shortest path back to its starting point. Therefore, it is reasonable that the sub-Riemannian vortex geodesics derived by the algebraic approach coincides with the point vortex equilibrium.

%\begin{itemize}
% \item Assumption: Point vortex equilibria are the shortest paths, i.e. geodesics 
% \item Motion of three point vortices on a sphere by Kidambi, Newton, Physica D:
%    For fixed equilibria, the vortices must lie on great circles (geodesics)
% \item Variationsprinzip der Wirbeldynamik?
% \item Idee: Minimal Energy $\longrightarrow$ Equilibrium $\longrightarrow$ Geodesic
% \item Kelvin, life, Labours and Legacy by R.Flood et al:
% The development proceeds to obtain Ã¢ÂÂwhat posterity regards as the single most
%important aspect of the treatiseÃ¢ÂÂ This aspect involved basing their dynamics on
%a variational principleÃ¢ÂÂthe principle of least action. This action they defined as
%twice the time integral of the kinetic energy, and the principle of least action is:
%Of all the different sets of paths along which a conservative system may be guided to
%move from one configuration to another, with the sum of its potential and kinetic energies
%equal to a given constant, that one for which the action is the least is such that the
%system will require only to be started with the proper General velocities, to move along
%it unguid
%\item Mathematical aspects of vortex dynamics (Campbell S.202), Nonlinear stability of fluid and plasma e
%quilibria (D.Holm) and On vortex lattices (Tkachenko): vortices have minimal energy on triangular lattices!
%I.e. 3 Point vortex equilibria have minimal energy! 

%--------------------------------------------------------

\section{How can 3D vortex geodesics be derived?} 
\label{sec:3-6}

In this section we will outline the concept to  apply (3,6)-sub-Riemannnian geometry to the Vortex-Heisenberg algebra vh(3). 
The question arises what \textit{are} geodesics in three-dimensional flows physically? 
The idealized two-dimensional point vortex systems can be seen as the intersection of a plane with (straight) vortex lines. Therefore, regarding three-dimensional vortex flows, we suggest to relate vortex geodesics to the motion of vortex lines. A vortex line is a material line that is composed of many Lagrangian particles with infinitesimal rotation.

%The Vortex-Heisenberg Lie algebra vh(3) representing the dynamics of three dimensional vortex flows has six dimensions, because it is generated by three components of the momentum $\mathbf{P} = (P_x,P_y,P_z)$ and three components of the total flux of vorticity $\mathbf{Z} = (Z_x,Z_y,Z_z)$. 
%The bracket relations with respect to the helicity $h_V$ read as
%\begin{equation}
%\{P_i,h_V,P_j\} = \epsilon_{ijk} Z_k. %, \quad  \{Z_i,h_V,P_j\} =  \{Z_i,h_V,Z_j\} = 0.
%\label{3DWirbbracket}
%\end{equation}
The Vortex Heisenberg Lie algebra vh(3) is endowed with a rank $n=3$ distribution. Therefore, we can apply (3,6)-sub-Riemannnian geometry to vh(3) aiming for the derivation of 3D vortex geodesics.  The fist investigations of (3,6)-sub-Riemannian geometry can be ascribed to \citet{Myasnichenko2002}.
%. We apply (3,6)-sub-Riemannnian geometry to our representation of fluid dynamics because of the following reasons:. %, givechapter \ref{chap:ContinNam}. 
%Thus, we will consider a sub-Riemannian space for the search for vortex geodesics.
Here, we will proceed analogously to the previously discussed (2,3)-vortex geodesics in section \ref{sec:2.3allg} after \cite{Perez2006}.

%Mathematically, the Lie algebra vh(3) is nilpotent and therefore, it yields a natural sub-Riemannian structure. 
We have discussed that sub-Rieman\-nian geometry is used to find geodesics for systems, where the motion is restricted.
Regarding vortex dynamics, the motion is restricted by physical conservation laws  such as the conservation of the flux of vorticity.
Moreover, the conservation of the linear momenta, energy and helicity play a crucial role restrict the motion of the vortices.
We note that we do not take the conservation of energy into account, but the vortex-related conserved quantities to determine the geodesics. 

The geodesics are given by the following differential equation (see \eqref{eq:normed_curve})
\begin{equation}
 \frac{d\mathbf{x}_g}{dt}  = (\mathbf{\dot{h}}, \dot{ \underline{\mathbf{H}}}) = ( \underline{\mathbf{H}}\mathbf{h}, \mathbf{0})
 \label{eq:geodH3W}. 
\end{equation}
Thus, we first have to determine thepair $(\mathbf{h}, \underline{\mathbf{H}})$ (see \eqref{eq:pair}). 
The tensor $\underline{\mathbf{H}} $ is given by the Nambu bracket of the linear momenta $\mathbf{P} = (P_x,P_y,P_z)$: 
 \begin{equation}
\underline{\mathbf{H}}
=  \begin{pmatrix} 
 \{P_x,h_V,P_x \} & \{P_x,h_V,P_y\} & \{P_x,h_V,P_z\} \\
 \{P_y,h_V,P_x\} & \{P_y,h_V,P_y\} & \{P_y,h_V,P_z\} \\
\{P_z,h_V,P_x\} & \{P_z,h_V,P_y\}  & \{P_z,h_V,P_z\} 
\end{pmatrix}, 
\end{equation}
where the Nambu bracket of  the linear momenta with respect to the helicity results in the total flux of vorticity.
%We now discretize the total flux of vorticity $\mathbf{Z}$ and the momenta $P_x, P_y$ and $P_z$  by assuming a constant volume equal to one 
%$V_i = 1$ for all $i$ volumes. 

The Vortex-Heisenberg algebra is derived for continuous vortex dynamics, but we aim for discrete 3D vortex geodesics. 
Therefore, we discretize the global fields of the momentum $\mathbf{P}= (P_x,P_y,P_z)$ that is given by the integral over the velocity $\mathbf{v}$ and the total flux of vorticity $\mathbf{Z}= (Z_x,Z_y,Z_z)$ that is defined by the integral over the vorticity vector $\boldsymbol{\xi}$. 
Approximating the integrals by $n$ summands
\begin{equation}
\mathbf{Z}  = \int_V d\tau \  \boldsymbol{\xi}
 \approx  \sum_{i=m}^n  \Delta V_m\ \boldsymbol{\xi}_m
\end{equation}
and
\begin{equation}
\mathcal{\mathbf{P}}  =  \frac{1}{2} \int_V d\tau \ (\mathbf{r} \times \boldsymbol{\xi} ) \approx \sum_{i=m}^n \Delta V_m \ (\mathbf{r} \times \boldsymbol{\xi} )_m
%=  \int_V d\tau \ \mathbf{v} \approx \sum_{i=m}^n \Delta V_m \ \mathbf{v}_m,
\end{equation} 
we divide the flow field into $n$ pieces, each with unit volume $\Delta V_i$:
Then, the discretized pair $(\mathbf{h}, \underline{\mathbf{H}})$ is given by
 the anti-symmetric rotational tensor and the vector of the relative momenta
 \begin{equation}
\underline{\mathbf{H}}
=
\begin{pmatrix}
0 & Z_z & -Z_y \\
-Z_z & 0 & Z_x \\
Z_y & -Z_x & 0 
\end{pmatrix},
\quad 
\mathbf{h} = 
\begin{pmatrix}
P_x\\
P_y\\
P_z
\end{pmatrix}.
\end{equation}
In two-dimensions, we considered the momenta of each point vortex separately, and the common total circulation of the whole point vortex system. Here, we regard each vortex line separately and use the index ${(i)}$ to mark the $i$-the vortex line.
To solve \eqref{eq:geodH3W} for this system we apply Lagrange-Sylvester formula that is the spectral formula for $\exp (t \underline{\mathbf{H}}) $ \eqref{eq:expfun}.
Thus, we first need to calculate the eigenspaces and projectors. 
Therefore, we apply the eigenwert equation and obtain for the $i$-th  vortex line:
\begin{equation}
\begin{split}
\det(\underline{\mathbf{H}}^{(i)}-&\kappa \underline{\mathbf{E}}) 
 =  \det \begin{pmatrix}
-\kappa & Z_z^{(i)} & -Z_y^{(i)} \\
-Z_z^{(i)} & -\kappa & Z_x^{(i)} \\
Z_y^{(i)} & -Z_x^{(i)} & -\kappa
\end{pmatrix}  \\
 = & -\kappa^3 + Z_z^{(i)} Z_x^{(i)}Z_y^{(i)}- Z_y^{(i)}(-Z_z^{(i)})(-Z_x^{(i)})
- Z_y^{(i)}(-\kappa)(-Z_y^{(i)}) \\
& -(Z_x^{(i)})Z_x^{(i)}(-\kappa)
+\kappa(-Z_z^{(i)})Z_z^{(i)} 
=-\kappa^3-\kappa({Z_y^{(i)}}^2+{Z_x^{(i)}}^2+{Z_z^{(i)}}^2) \\
 = & -\kappa (\kappa^2+({Z_y^{(i)}}^2+{Z_x^{(i)}}^2+{Z_z^{(i)}}^2)
= -\kappa (\kappa^2+\vert {\mathbf{Z}^{(i)}}^2 \vert^2) .
\end{split}
\end{equation}
Thus, 
\begin{equation}
\det(\underline{\mathbf{H}}^{(i)}-\kappa \underline{\mathbf{E}})   = 0  \Longleftrightarrow \ \kappa = 0 \quad \text{or} \quad \kappa = i \vert \mathbf{Z}^{(i)} \vert.
\end{equation}
We summarize the set of eigenvalues
\begin{equation}
\{\kappa_1, \kappa_2,\kappa_3 \} 
= \{i \lambda, - i \lambda,0 \} 
= \{i \vert \mathbf{Z}^{(i)} \vert, - i \vert \mathbf{Z} ^{(i)}\vert,0 \} , \quad
\Rightarrow \lambda = \vert \mathbf{Z} ^{(i)}\vert ,
\end{equation}
where $\underline{\mathbf{H}}^{(i)}$ and $\lambda$ satisfy the condition $\underline{\mathbf{H}}^{(i)}({\underline{\mathbf{H}}^{(i)}}^2+ \lambda^2 \underline{\mathbf{E}}) =0$ as proposed in \citet{Perez2006}. 
%\color{blue}(Hat det$ \underline{\mathbf{H}} = 0$ eine Bedeutung?) \color{black}
We now determine the projectors $\underline{\boldsymbol{\pi}}_0^{(i)}$ and $\underline{\boldsymbol{\pi}}_1^{(i)}$:
\begin{equation}
\begin{split}
& \underline{\boldsymbol{\pi}}_0^{(i)} = \frac{1}{\lambda^2} ({\underline{\mathbf{H}}^{(i)}} ^2+\lambda^2\underline{\mathbf{E}}  ) =  \frac{1}{\vert {\mathbf{Z}^{(i)}} \vert^2}  \cdot \\ 
&
\begin{pmatrix}
-{Z_z^{(i)}}^2-{Z_y^{(i)}}^2-\vert \mathbf{Z} ^{(i)}\vert^2 & Z_x^{(i)} Z_y ^{(i)}& Z_y^{(i)} Z_z^{(i)} \\
Z_x^{(i)} Z_y^{(i)} & -{Z_z^{(i)}}^2-{Z_x^{(i)}}^2-\vert \mathbf{Z}^{(i)} \vert^2 &Z_y ^{(i)}Z_z^{(i)} \\
Z_x^{(i)} Z_z^{(i)} & Z_y^{(i)} Z_z^{(i)} & -{Z_y^{(i)}}^2- {Z_x^{(i)}}^2 -\vert \mathbf{Z}^{(i)} \vert^2
\end{pmatrix} 
\end{split} 
 \end{equation}
 and
\begin{equation}
\begin{split}
\underline{\boldsymbol{\pi}}_1^{(i)} &= \frac{1}{2\lambda^2} \underline{\mathbf{H}}^{(i)} (\underline{\mathbf{H}}^{(i)} +i\lambda \underline{\mathbf{E}} ) =  \frac{1}{2\vert \mathbf{Z}^{(i)} \vert^2} \cdot \\
& \begin{pmatrix}
-{Z_z^{(i)}}^2-{Z_y^{(i)}}^2-i\vert \mathbf{Z} ^{(i)}\vert & Z_x^{(i)} Z_y^{(i)} & Z_y^{(i)} Z_z^{(i)} \\
Z_x^{(i)} Z_y^{(i)} & -{Z_z^{(i)}}^2-{Z_x^{(i)}}^2- i \vert \mathbf{Z} \vert &Z_y^{(i)} Z_z^{(i)} \\
Z_x^{(i)} Z_z^{(i)} & Z_y^{(i)} Z_z^{(i)} & -{Z_y^{(i)}}^2- {Z_x^{(i)}}^2 -i \vert \mathbf{Z}^{(i)} \vert
\end{pmatrix}
\end{split}
 \end{equation} 
The eigenvector for the zero eigenvalue reads as
\begin{equation}
\mathbf{v}_0^{(i)} = 
 \frac{1}{\vert \lambda \vert }
\cdot \begin{pmatrix}
H_{23}^{(i)} \\ -H_{13}^{(i)} \\ H_{12} ^{(i)}
\end{pmatrix}
=
\frac{1}{\vert \mathbf{Z}^{(i)} \vert }
\cdot 
\begin{pmatrix}
Z_z^{(i)} \\ Z_x ^{(i)}\\ Z_y ^{(i)}
\end{pmatrix}.
\end{equation}
Thus, the basis vector $\boldsymbol{\gamma}_0^{(i)}$  for the $i$-th vortex line is given by:
 \begin{equation}
 \begin{split}
\boldsymbol{\gamma}_0^{(i)} & = 
\frac{1}{\lambda} (H_1^{(i)} H_{23}^{(i)} - H_2^{(i)} H_{13} ^{(i)}+ H_3^{(i)} H_{12}^{(i)}) \mathbf{v}_0^{(i)}\\
  & = \frac{1}{\vert \mathbf{Z}^{(i)}\vert^2 }(P_x^{(i)} Z_x^{(i)}+ P_y^{(i)} Z_y^{(i)} + P_z^{(i)} Z_y^{(i)}) 
  \begin{pmatrix}
  Z_x^{(i)} \\ Z_y^{(i)} \\ Z_z ^{(i)}
  \end{pmatrix} \\
   & =  \frac{1}{ \vert \mathbf{Z}^{(i)}\vert^2 } 
   (\mathbf{P}^{(i)}\cdot \mathbf{Z}^{(i)})\mathbf{Z}^{(i)}
  \end{split}
 \end{equation}
and the basis vectors $\boldsymbol{\beta}^{(i)}$  and $\boldsymbol{\alpha}^{(i)}$ read as:
\begin{equation}
\boldsymbol{\beta}^{(i)} 
= \frac{1}{2} \left( \mathbf{P}^{(i)} - \boldsymbol{\gamma}_0 \right) 
= \frac{1}{2} \left( \mathbf{P}^{(i)}-\frac{1}{\vert \mathbf{Z}^{(i)}\vert^2 } 
   (\mathbf{P}^{(i)}\cdot \mathbf{Z}^{(i)})\mathbf{Z}^{(i)}\right)
\end{equation}
and
\begin{equation}
\begin{split}
\boldsymbol{\alpha}^{(i)} &= - \frac{1}{\lambda} \underline{\mathbf{H}}^{(i)}\boldsymbol{\beta}^{(i)} \\
&= \frac{1}{- \vert \mathbf{Z}^{(i)} \vert }
\begin{pmatrix}
0 & Z_z^{(i)} & -Z_y^{(i)} \\
-Z_z^{(i)} & 0 & Z_x^{(i)} \\
Z_y^{(i)} & Z_x^{(i)} & 0 \\
\end{pmatrix}
\cdot \frac{1}{2} \left( \mathbf{P}^{(i)}- \frac{1}{i \vert \mathbf{Z}^{(i)}\vert^2 } 
   (\mathbf{P}^{(i)}\cdot \mathbf{Z}^{(i)})\mathbf{Z}^{(i)}\right) \\
& = \frac{1}{2 \vert \mathbf{Z}^{(i)}\vert} \mathbf{P}^{(i)} \times \mathbf{Z} ^{(i)}
+\frac{1}{2 \vert \mathbf{Z}^{(i)}\vert ^3} ( \mathbf{P} ^{(i)}\cdot \mathbf{Z} ^{(i)}) \ (\mathbf{Z} ^{(i)}\times \mathbf{Z}^{(i)} ) \\
& = \frac{1}{2 \vert \mathbf{Z}^{(i)}\vert} \mathbf{P} ^{(i)}\times \mathbf{Z}^{(i)}
\end{split}.
\end{equation}
Now, we have derived a basis $\{\boldsymbol{\alpha}^{(i)},\boldsymbol{\beta}^{(i)},\boldsymbol{\gamma}_0^{(i)}\}$ for each discrete vortex line. 
And we can formulate the geodesics equation \eqref{eq:TrajecSubrgen} with respect to the basis leading to an equation for geodesics for three-dimensional incompressible, inviscid flows.% \color{red}vollstaendig?\color{black}:

\begin{figure}
\includegraphics[scale=.7]{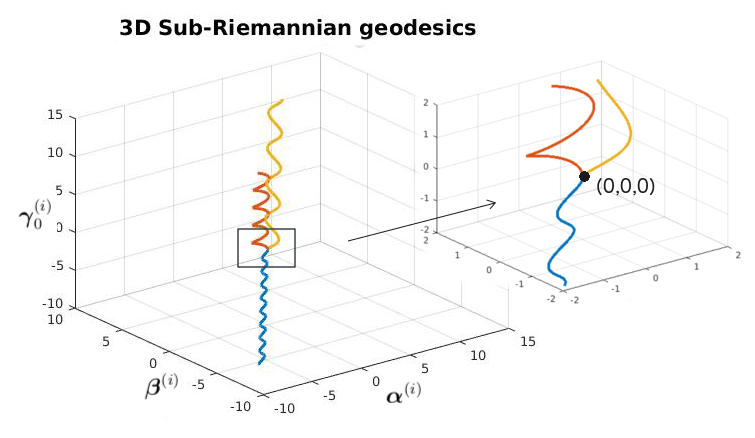}
\caption{\small {Three examples of 3D vortex geodesics derived by (3,6)-Sub-Riemannian geodesics are shown. Because of the initial condition, all three vortex lines intersect in the origin. 
}}
\label{fig:3DSubrie}
\end{figure}

Then, the 3D geodesics are given by:
\begin{equation}
\boxed{
\begin{split}
\mathbf{x}_g^{(i)}  &= \frac{1}{ 2 \vert \mathbf{Z}^{(i)} \vert ^2}
\left[ \cos(\vert \mathbf{Z} ^{(i)}\vert t -1) \right]
(\mathbf{P}\times \mathbf{Z}^{(i)}) \\
&+\frac{1}{2 \vert \mathbf{Z}^{(i)} \vert } \sin(\vert \mathbf{Z}^{(i)} \vert t ) \ 
\left(\mathbf{P} ^{(i)}- \frac{1}{\vert \mathbf{Z}^{(i)} \vert ^2} (\mathbf{P}^{(i)}\cdot \mathbf{Z}^{(i)}) \mathbf{Z}^{(i)} \right) \\
&+ t \  
\frac{1}{ \vert \mathbf{Z}^{(i)}\vert^2 } 
   (\mathbf{P}^{(i)}\cdot \mathbf{Z}^{(i)})\mathbf{Z}^{(i)}
\end{split}
}
\end{equation}
One example is shown in fig. \ref{fig:3DSubrie}. Here we chose the following local coordinates and vorticity vectors  $\mathbf{x}_1 = (1,0,1)$, $\boldsymbol{\xi}_1 = (2,1,5)$, $\mathbf{x}_2 = (1,-1,0)$, $\boldsymbol{\xi}_2 = (1,3,1)$ and $\mathbf{x}_3 = (0,4,0)$, $\boldsymbol{\xi}_3 = (2,1,5)$. 
%As indicated by \citet{Perez2006} 
We see that all lines are helices that start from the origin, which is due to the initial condition. The helical structure is discussed by 
\citet{Perez2006} for general nilpotent algebras.

In the last section we explored two-dimensional flows, where vortex geodesics can be interpreted as motion of Lagrangian particles. 
Here, we derived geodesics for three-dimensional vortex flows, where we could think of \textit{frozen vortex lines} as three-dimensional discrete vortices indicated by the colored lines in fig. \ref{fig:3DSubrie}. 
Analogously to the 2D-vortex geodesics, we might interpret these helical curves as discrete, 3D \textit{equilibrium} solutions of vortex flows. 
From meteorological perspective, examples of vortex lines are the centerlines of tropical cyclones.

We conclude that 3D vortex geodesics for incompressible, inviscid flows can be derived from the nilpotent algebra for vortex dynamics.
They are geo\-metrically given by helices. In three dimensions it is hard to compare this helical motion with previous studies, since we could not find an idealized, analogous concept for three dimensions, as we did in two dimensions, where we could compare the geodesics with point vortex equilibria. But, we recall that the Vortex-Heisenberg algebra is based on the conservation laws of the helicity, the total flux of vorticity and the 3D linear momentum. All of them contain the vorticity as non-rigid rotational vortex-quantity. 
Therefore, the helical paths of the vortices seem to be natural, but we will investigate 3D vortex geodesics more in detail in future studies. 
Here, we only outline a concept to derive 3D vortex geodesics starting from the Vortex-Heisenberg algebra.

%The here derived three-dimensional vortex geodesics might be compared with the concept of knots and braids. Already \citet{Moffatt1969} related idealized vortex lines, represented as knots, to the helicity, see also \cite{Ricca1992}, \citet{Moffatt2014}. We will examine these relations to vortex geodesics in further studies.

%Here, the initial total flux of vorticity %$Z_1 = Z_2 = Z_3 =1$ and the initial coordinates $(x_1,y_1,z_1) = (1,0,1)$, $(x_2,y_2,z_2) = (1,-1,0)$ and $(x_3,y_3,z_3) = (0,1,-1)$ were chosen.
%\color{blue} (Zusaetzlicher Term fuer ungerade n eingebaut?) \color{black}

\section{Summary}
\label{sec:summary}

The physical conservation laws that restrict the motion of fluid flows as well as the mathematical, nilpotent structure of the Vortex-Heisenberg algebras vh(2) and vh(3) motivated us to apply sub-Riemannian geometry to find point vortex geodesics.
We have also introduced a concept to derive of vortex geodesics for three-dimensional flows. The question arose, what are vortex geodesics in three dimensions? The answer could be: frozen vortex lines. 

However, by applying sub-Riemannian geometry to the Vortex-Heisen\-berg algebras that are based on the Nambu formulation for incompressible fluids we explore vortex dynamics from a reference system, where we \textit{sit on the vortex} and move with the vortex. In two spatial dimensions, we can imagine to move with a point vortex, whereas in three dimensions we consider vortex tubes, or vortex lines that are vortex tubes with infinitesimal radii.

So far, the Riemannian view has been used for the investigation of extremal principles for hydrodynamic systems, where mostly the energy is considered to derive variational principles for fluid dynamics. 
This is due to the system of equations, where the extremal principles are derived from. For incompressible hydrodynamical systems, most authors consider the Euler equations which describe the time evolution of the velocity. 
Then, it is an obvious choice to explore further problems in terms of the Hamiltonian structure determined by the kinetic energy. Some authors also regard the enstrophy, but the helicity rarely is taken into account. 
Our study is based on the Helmholtz equation that is obtained by the rotation of the Euler equations and describes the evolution of the vorticity. 
The Nambu bracket is directly based on the Helmholtz vorticity equation, which leads to Lie algebras that are based on vortex-related conservation laws.
These conservation laws determine and at the same time restrict the vortex motions.
Thus, we applied sub-Riemannnian geometry to vortex dynamics. 
To the best of our knowledge, there have been no investigations of the derivation of vortex geodesics in terms of sub-Riemannian geometry before. 

We applied the algorithm of \citet{Perez2006} to the two- and three-dimensional Vortex-Heisenberg algebras vh(2) and vh(3). 
For 2D as well as for 3D the phase space is constructed such that the sub-Riemannian trajectories pass the origin. But in 2D, a simple translation of the trajectories shows the congruence of the derived vortex geodesics and a planar point vortex motion, where the point vortices form an equilibrium. 
Why are point vortex equilibria and sub-Riemannian geodesics congruent?  
To answer this question, we consider initial local coordinates of one point vortex which is part of a two-or three point vortex system. 
This vortex starts moving and wants to arrive its initial position as soon as possible again. Its motion is restricted by fluid dynamical conservation laws that contain the rotational aspect, i.e. the circulation.
There are three kinds of idealized point vortex motion: (i) Collapse and expanding motion. In this case, the vortices do not arrive at their initial points again. (i) Arbitrary periodic motion: Here, the vortices rotate around the center of circulation several times until they reach their initial position again. But if the three vortices form (iii) an equilibrium with non-vanishing total circulation, 
they only rotate once around the center of circulation until they reach the initial coordinates again. Therefore, we think that a vortex takes the shortest path if and only if it is part of a equilibrium constellation. This explains why the trajectories of a point vortex equilibrium and sub-Riemannnian vortex geodesics are congruent. 

If the total sum of circulation is equal to zero, the point vortex system equilibrium translates. 
This vortex motion also coincides with the sub-Riemannian view, because in this case, we obtain an abnormal extremal which mathematically leads to a straight line as geodesic.
It would be interesting to investigate sub-Riemannian geometry for $N$-point vortex constellations, where $N\geq 4$. We hypothesize that such geodesics would also be congruent to relative equilibria.

In three dimensions, we find that the vortex geodesics are geometrically given by helices. 
Regarding the conserved quantities for three dimensional vortex flows, the helicity, the total flux of vorticity and the linear momentum, the helical paths seem to be natural, because all conservation laws contain the vorticity as rotational part. 
The geodesics could be thought of as \textit{frozen vortex lines} as three-dimensional discrete vortex structures. 
Analogously to the 2D-vortex geodesics, we might interpret these helical curves as discrete, 3D \textit{equilibrium} solutions of vortex flows. 
We will examine three-dimensional vortex geodesics more in detail in future studies. 

With this approach for the derivation of vortex dynamics we have shown that the algebraic view on vortex dynamics provides the possibilities for alternative descriptions and views on vortex dynamics.

\section*{Acknowledgements}
 This research has been funded by Deutsche Forschungsgemeinschaft (DFG) through grant CRC 1114 'Scaling Cascades in Complex Systems, Project Number 235221301, Projects A01 'Coupling a multiscale stochastic precipitation model to large scale atmospheric flow dynamics'.

 %\ref{fig:Subriemann:sketch}. 
%%\twocolumn

%\bibliographystyle{unsrtnat}
%%%\bibliographystyle{abbrvnat} 
\bibliography{references}

\begin{thebibliography}{}

\bibitem[Agrachev et~al., 2016]{Barilari2012}
Agrachev, A.~A., Barilari, D., and Boscain, U. (2016).
\newblock Introduction to riemannian and sub-riemannian geometry.
\newblock Accessed: 2018-02-25.

\bibitem[Aref, 1979]{Aref1979}
Aref, H. (1979).
\newblock Motion of three vortices.
\newblock {\em Physics of Fluids (1958-1988)}, 22(3):393--400.

\bibitem[Aref, 2007]{Aref2007}
Aref, H. (2007).
\newblock Point vortex dynamics: A classical mathematics playground.
\newblock {\em J. math. phys.}, 48(6):065401.

\bibitem[Arnold, 1969a]{Arnold1969}
Arnold, V.~I. (1969a).
\newblock {Hamiltonian nature of the Euler equations in the dynamics of a rigid
  body and of an ideal fluid}.
\newblock In {\em Vladimir I. Arnold-Collected Works}, pages 175--178.
  Springer.

\bibitem[Arnold, 1969b]{Arnold1969one}
Arnold, V.~I. (1969b).
\newblock {On one-dimensional cohomology of the Lie algebra of divergence-free
  vector fields and on rotation numbers of dynamic systems}.
\newblock In {\em Vladimir I. Arnold-Collected Works}, pages 179--182.
  Springer.

\bibitem[Arnold and Khesin, 1992]{Arnold1992}
Arnold, V.~I. and Khesin, B.~A. (1992).
\newblock Topological methods in hydrodynamics.
\newblock {\em Annu. rev. fluid mech.}, 24(1):145--166.

\bibitem[Barilari et~al., 2016]{Barilari2016}
Barilari, D., Boscain, U., and Sigalotti, M. (2016).
\newblock {\em Geometry, analysis and dynamics on sub-Riemannian manifolds.
  Volume II}.
\newblock Z{\"u}rich: European Mathematical Society (EMS).

\bibitem[Berger and Gostiaux, 2012]{Berger2012}
Berger, M. and Gostiaux, B. (2012).
\newblock {\em Differential Geometry: Manifolds, Curves, and Surfaces}, volume
  115.
\newblock Springer Science \& Business Media.

\bibitem[Blackmore et~al., 2007]{Blackmore2007}
Blackmore, D., Ting, L., and Knio, O. (2007).
\newblock Studies of perturbed three vortex dynamics.
\newblock {\em J. math. phys.}, 48(6):065402.

\bibitem[Brockett, 1982]{Brockett1982}
Brockett, R.~W. (1982).
\newblock {\em Control theory and singular Riemannian geometry}, pages 11--27.
\newblock Springer.

\bibitem[Calin and Chang, 2009]{Calin2009}
Calin, O. and Chang, D.-C. (2009).
\newblock Sub-riemannian geometry.
\newblock {\em Cambridge University Press}, 24:134.

\bibitem[Chapman, 1978]{Chapman1978}
Chapman, D.~M. (1978).
\newblock Ideal vortex motion in two dimensions: symmetries and conservation
  laws.
\newblock {\em J. math. phys.}, 19(9):1988--1992.

\bibitem[Do~Carmo et~al., 2017]{DoCarmo2017}
Do~Carmo, M.~P., Fischer, G., Pinkall, U., and Reckziegel, H. (2017).
\newblock Differential geometry.
\newblock In {\em Mathematical Models}, pages 155--180. Springer.

\bibitem[Gr{\"o}bli, 1877]{Groebli1877}
Gr{\"o}bli, W. (1877).
\newblock {\em {Specielle Probleme {\"u}ber die Bewegung geradliniger
  paralleler Wirbelf{\"a}den}}.
\newblock Z{\"u}rcher und Furrer.

\bibitem[Helmholtz, 1858]{Helmholtz1858}
Helmholtz, H. (1858).
\newblock {{\"U}ber hydrodynamischer Gleichungen welche den Wirbelbewegungen
  entsprechen}.
\newblock {\em J. Reine Angew. Math}, 5:25--55.

\bibitem[Holm et~al., 1998]{Holm1998}
Holm, D.~D., Marsden, J.~E., and Ratiu, T.~S. (1998).
\newblock The euler--poincar{\'e} equations and semidirect products with
  applications to continuum theories.
\newblock {\em Advances in Mathematics}, 137(1):1--81.

\bibitem[Kirchhoff, 1876]{Kirchhoff1876}
Kirchhoff, G. (1876).
\newblock {\em {Vorlesungen {\"u}ber mathematische Physik I}}, volume~1.
\newblock Teubner.

\bibitem[K{\"u}hnel, 1999]{Kuhnel1999}
K{\"u}hnel, W. (1999).
\newblock {\em Differentialgeometrie}, volume 2003.
\newblock Springer.

\bibitem[Majda and Bertozzi, 2002]{Majda2002}
Majda, A.~J. and Bertozzi, A.~L. (2002).
\newblock {\em Vorticity and incompressible flow}, volume~27.
\newblock Cambridge University Press.

\bibitem[Makhaldiani, 1998]{Makhaldiani1998}
Makhaldiani, N. (1998).
\newblock {The system of three vortexes of two dimensional ideal hydrodynamics
  as a new example of the (integrable) Nambu-Poisson mechanics}.
\newblock {\em arXiv preprint solv-int/9804002}.

\bibitem[Marsden and Weinstein, 1983]{Marsden1983}
Marsden, J. and Weinstein, A. (1983).
\newblock {Coadjoint orbits, vortices, and Clebsch variables for incompressible
  fluids}.
\newblock {\em Physica D: Nonlinear Phenomena}, 7(1):305--323.

\bibitem[Monroy-P{\'e}rez and Anzaldo-Meneses, 1999]{Monroy1999}
Monroy-P{\'e}rez, F. and Anzaldo-Meneses, A. (1999).
\newblock {Optimal control on the Heisenberg group}.
\newblock {\em Journal of dynamical and control systems}, 5(4):473--499.

\bibitem[Monroy-P{\'e}rez and Anzaldo-Meneses, 2006]{Perez2006}
Monroy-P{\'e}rez, F. and Anzaldo-Meneses, A. (2006).
\newblock The step-2 nilpotent (n,n(n+1)/2) sub-riemannian geometry.
\newblock {\em Journal of dynamical and control systems}, 12(2):185--216.

\bibitem[Montgomery, 2006]{Montgomery2006}
Montgomery, R. (2006).
\newblock {\em A tour of subriemannian geometries, their geodesics and
  applications}.
\newblock Number~91. American Mathematical Soc.

\bibitem[M{\"u}ller, 2018]{mueller2018algebraic}
M{\"u}ller, A. (2018).
\newblock {\em On algebraic and geometric aspects of fluid dynamics: New
  perspectives based on Nambu mechanics and its applications to atmospheric
  dynamics}.
\newblock PhD Thesis, Freie Universitaet Berlin (Germany).

\bibitem[M\"{u}ller and N{\'e}vir, 2014]{Mueller2014}
M\"{u}ller, A. and N{\'e}vir, P. (2014).
\newblock A geometric application of {N}ambu mechanics: the motion of three
  point vortices in the plane.
\newblock {\em Journal of Physics A: Mathematical and Theoretical},
  47(10):105201.

\bibitem[M{\"u}ller and N{\'e}vir, 2021]{muller2021algebra}
M{\"u}ller, A. and N{\'e}vir, P. (2021).
\newblock On the algebra and groups of incompressible vortex dynamics.
\newblock {\em Journal of Physics A: Mathematical and Theoretical},
  54(30):305203.

\bibitem[Myasnichenko, 2002]{Myasnichenko2002}
Myasnichenko, O. (2002).
\newblock Nilpotent (3,6)-sub-riemannian problem.
\newblock {\em Journal of dynamical and control systems}, 8(4):573--597.

\bibitem[N\'{e}vir, 1998]{Nevir1998}
N\'{e}vir, P. (1998).
\newblock {\em Die {N}ambu-{F}elddarstellungen der {H}ydro-{T}hermodynamik und
  ihre {B}edeutung f\"ur die dynamische {M}eteorologie}.
\newblock Habilitation dissertation, FU Berlin.

\bibitem[N{\'e}vir and Blender, 1993]{Nevir1993}
N{\'e}vir, P. and Blender, R. (1993).
\newblock {A Nambu representation of incompressible hydrodynamics using
  helicity and enstrophy}.
\newblock {\em J phys. A -- math. gen.}, 26(22):L1189.

\bibitem[Newton, 2001]{Newton2001}
Newton, P.~K. (2001).
\newblock {\em The N-vortex problem: analytical techniques}, volume 145.
\newblock Springer Science \& Business Media.

\bibitem[Novikov, 1975]{Novikov1975}
Novikov, E.~A. (1975).
\newblock Dynamics and statistics of a system of vortices.
\newblock {\em Zh. Eksp. Teor. Fiz}, 68:1868--1882.

\bibitem[Obukhov et~al., 1984]{Obukhov1984}
Obukhov, A.~M., Kurgansky, M.~V., and Tatarskaya, M.~S. (1984).
\newblock Dynamic conditions for the origin of drought and other large-scale
  weather anomalies (russian).
\newblock {\em Meteorologiia i Gidrologiia}, pages 5--13.

\bibitem[O'Neill, 2006]{ONeill2006}
O'Neill, B. (2006).
\newblock {\em Elementary differential geometry}.
\newblock Academic press.

\bibitem[Percacci, 2017]{Percacci2017}
Percacci, R. (2017).
\newblock Lecture notes on lie groups.
\newblock Accessed: 2017-12-27.

\bibitem[Salmon, 1982]{Salmon1982}
Salmon, R. (1982).
\newblock {Hamilton's principle and Ertel's theorem}.
\newblock {\em AIP Conference Proceedings}, 88(1):127--135.

\bibitem[Salmon, 1988]{Salmon1988}
Salmon, R. (1988).
\newblock Hamiltonian fluid mechanics.
\newblock {\em Annu. rev. fluid mech.}, 20(1):225--256.

\bibitem[Shepherd, 1990]{Shepherd1990}
Shepherd, T.~G. (1990).
\newblock {\em Symmetries, conservation laws, and Hamiltonian structure in
  geophysical fluid dynamics}, volume~32, pages 287--338.
\newblock Elsevier.

\bibitem[Synge, 1949]{Synge1949}
Synge, J.~L. (1949).
\newblock On the motion of three vortices.
\newblock {\em Can. J. Math}, 1(3):257--270.

\end{thebibliography}

\end{document}